\documentclass[a4paper,fleqn,usenatbib, hyperref]{mnras}

\usepackage[T1]{fontenc}
\usepackage{ae,aecompl}
\usepackage{hyperref}
\usepackage{graphicx}
\usepackage{mathrsfs}
\usepackage{amsmath}
\usepackage{amssymb}
\usepackage{verbatim}
\usepackage{multirow}
\usepackage{subfig}
\usepackage{mathtools}
\usepackage[noabbrev]{cleveref}
\def\apjl{ApJ}
\def\araa{ARA\&A}             
\def\apj{ApJ}                 
\def\apjl{ApJ}                
\def\apjs{ApJS}               
\def\ao{Appl.~Opt.}           
\def\apss{Ap\&SS}             
\def\aap{A\&A}                
\def\aaps{A\&AS}              
\def\mnras{MNRAS}             
\def\qjras{QJRAS}             
\def\nat{Nature}              
\def\iaucirc{IAU~Circ.}       

\def\physrep{Phys.~Rep.}   

\def \chandra {{\em Chandra}}
\def \xmm {{\em XMM-Newton}}
\def \rxte{{Rossi X-ray Timing Explorer}}

\def \saxj{SAX J1808.4-3658}
\def \igrj{IGR J00291+5934}
\def \xt{XTE J1751-305}
\def \igj{IGR J17511-3057}

\def \nustar{{\em NuSTAR}}

\title[Timing of the AMXP \saxj{}]{On the timing properties of \saxj{} during its 2015 outburst}
\author[Sanna et al. ]{A. Sanna$^{1}$\thanks{E-mail:
    andrea.sanna@dsf.unica.it}, T. Di Salvo$^{2}$, L. Burderi$^{1}$, A. Riggio$^{1}$, F. Pintore$^{3}$, A. F. Gambino$^{2}$ \newauthor
     R. Iaria$^{2}$, M. Tailo$^{1,4}$, F. Scarano$^{1}$, A. Papitto$^{4}$\\
$^{1}$Dipartimento di Fisica, Universit\`a degli Studi di Cagliari, SP Monserrato-Sestu km 0.7, 09042 Monserrato, Italy\\
$^{2}$Universit\`a degli Studi di Palermo, Dipartimento di Fisica e Chimica, via Archirafi 36, 90123 Palermo, Italy\\
$^{3}$INAF-Istituto di Astrofisica Spaziale e Fisica Cosmica - Milano, via E. Bassini 15, I-20133 Milano, Italy\\
$^{4}$INAF-Osservatorio Astronomico di Roma, Via di Frascati 33, I-00044, Monteporzio Catone (Roma), Italy}
\begin{document}

\date{Accepted -. Received -; in original form -}

\pagerange{\pageref{firstpage}$-$\pageref{lastpage}} \pubyear{2015}

\maketitle

\label{firstpage}

\begin{abstract}
We present a timing analysis of the 2015 outburst of the accreting millisecond X-ray pulsar \saxj{}, using non-simultaneous \xmm{} and \nustar{} observations. We estimate the pulsar spin frequency and update the system orbital solution. Combining the average spin frequency from the previous observed, we confirm the long-term spin down at an average rate $\dot{\nu}_{\text{SD}}=1.5(2)\times 10^{-15}$ Hz s$^{-1}$. We also discuss possible corrections to the spin down rate accounting for mass accretion onto the compact object when the system is X-ray active. Finally, combining the updated ephemerides with those of the previous outbursts, we find a long-term orbital evolution compatible with a binary expansion at a mean rate $\dot{P}_{orb}=3.6(4)\times 10^{-12}$ s s$^{-1}$, in agreement with previously reported values. This fast evolution is incompatible with an evolution driven by angular momentum losses caused by gravitational radiation under the hypothesis of conservative mass transfer. We discuss the observed orbital expansion in terms of non-conservative mass transfer and gravitational quadrupole coupling mechanism. We find that the latter can explain, under certain conditions, small fluctuations (of the order of few seconds) of the orbital period around a global parabolic trend. At the same time, a non-conservative mass transfer is required to explain the observed fast orbital evolution, which likely reflects ejection of a large fraction of mass from the inner Lagrangian point caused by the irradiation of the donor by the magneto-dipole rotator during quiescence ({\it radio-ejection} model). This strong outflow may power tidal dissipation in the companion star and be responsible of the gravitational quadrupole change oscillations.

\end{abstract}


\section{Introduction}

\saxj{} is a neutron star (NS) low-mass X-ray binary (LMXB) observed for the first time in 1996 by the \textit{BeppoSAX} satellite \citep{in-t-Zand1998a}. The discovery of the coherent pulsation at roughly 2.5 milliseconds \citep{Wijnands1998a, Chakrabarty1998a} made this source the first observed accreting millisecond X-ray pulsar (AMXP). This class of objects are X-ray transients that generate X-ray coherent pulsation by accreting matter onto the NS polar caps. Matter from the companion star is firstly transferred via Roche-lobe overflow, and then interacts with the NS magnetosphere. AMXPs are characterised by long quiescence periods (years to decades) interrupted by short outburst phases (lasting weeks to months) during which the X-ray emission increases by orders of magnitude. Processes such as thermal-viscous instability in a thin disc might be responsible for the episodic accretion rate variations \citep[see e.g.,][]{Smak1982a, Meyer1982a}, although the outburst trigger mechanism is still highly debated. AMXPs have been proved to be the evolutionary link between  radio millisecond pulsars and accreting NS \citep[see e.g. IGR J18245$-$2452;][]{Papitto2013b}, confirming the so called \emph{recycling scenario} \citep[see e.g., ][]{Bhattacharya91}. 

Since its discovery, \saxj{} was observed in outburst seven times (with average recurrence times of roughly 2.5-3.5 years), including the most recent one started on April 2015 \citep{Sanna15d}. So far, \saxj{} has been the most observed of its class, making it the best candidate to explore the long-term properties of AMXPs. The source exhibited several thermonuclear X-ray bursts in each of the outbursts, some of which where classified as photospheric radius expansion X-ray bursts \citep{Galloway2006a, Galloway2008a}, most likely originating in a flash of pure helium layer created by stable burning hydrogen. Combining the X-ray burst from different outbursts, \citet{Galloway2006a} placed the source at a distance of $3.5\pm0.1$ kpc.

The secular evolution of the spin frequency, up to the 2011 outburst of the source, is compatible with a constant spin-down derivative of magnitude $\sim -10^{-15}$ Hz/s \citep{Patruno12a}, likely reflecting a magnetic-dipole torque acting during quiescence \citep[see also][]{Hartman08,Hartman09b}. These effect constraints the magnetic-dipole moment at the value $\mu\sim10^{26}$ G cm$^3$, corresponding to a surface magnetic field at the NS poles $B\sim2\times 10^{8}$ G \citep[assuming a NS radius of 10 km;][]{Hartman08}. Similar estimates have been inferred from spectral fitting, more precisely from the study of the broad Iron emission line \citep{Cackett09, Papitto09}, and from the modelling of the accretion disk \citep{Ibragimov2009a}. Combining the spin frequency and the source luminosity in quiescence, \citet{DiSalvo2003a} estimated a surface magnetic field in the range $(1-5)\times 10^{8}$ G. Moreover, the detection of a spin-down frequency derivative of $\sim-8\times 10^{-14}$ Hz s$^{-1}$ during the final stages of the 2002 outburst, suggested a NS magnetic field of $\sim 3.5\times 10^{8}$ G \citep{Burderi06}.

From the analysis of the orbital modulation of the persistent pulsation has been determined a binary orbital period of $\sim2$ hours \citep{Chakrabarty1998a}, around a companion star of mass likely in the range 0.04-0.14 M$_\odot$ \citep{Bildsten2001a, diSalvo08, Deloye2008a, Burderi09}. Combining the orbital ephemerides of the outbursts up to the 2005, \citet{diSalvo08} and \citet{Hartman08} observed an expansion of the orbital period at a rate of $\sim3.5\times 10^{-12}$ s s$^{-1}$. This result has been confirmed by \citet[][from the analysis of the \textit{Rossi X-ray Time Explorer} observations]{Hartman09b} and \citet[][from the analysis of the a single \xmm{} observation]{Burderi09} that extended the analysis up to the 2008 outburst. The authors found a consistent orbital period derivative, suggesting that the orbital evolution remained stable over the ten years baseline considered. The large orbital expansion rate has been interpreted by \citet{diSalvo08} and \citet{Burderi09} as the result of a highly non-conservative mass transfer where only a few percent of the transferred matter is accreted onto the NS. \citet{Hartman08, Hartman09b} suggested that, in analogy with a sample of observed ``black widow" millisecond radio pulsars \citep{Arzoumanian1994a, Doroshenko2001a}, the orbital period derivative of \saxj{} could have been the result of short-term interchanges of angular momentum between the companion star and the binary system. The latter hypothesis increased consensus after updating the long-term orbital evolution including the 2011 outburst of the source. \citet{Patruno12a} observed that the expansion of the orbital period of \saxj{} requires both a first and second orbital period derivative with values $\dot{P}_{orb}=3.5(2)\times 10^{-12}$ s s$^{-1}$ and $\ddot{P}_{orb}=1.65(35)\times 10^{-20}$ s s$^{-2}$, respectively. The authors proposed a mass quadrupole variation of the companion as the cause of the orbital period changes \citep{Applegate1994a}. \citet{Patruno2016a}, reporting a 30-ks \chandra{} observation of the source during the final stages of its 2015 outburst, highlighted that \saxj{} continues evolving with a large orbital period derivative. The authors explained the secular orbital behaviour of the source suggesting that either the NS ablates the donor causing a highly efficient mass-loss or the donor star undergoes quasi-cyclic variations caused by magnetically (or irradiation) driven mass-quadrupole changes.    

In this work, we carried out a coherent timing analysis of the 2015 outburst of \saxj{}, using not-simultaneous \xmm{} and \nustar{} data. We updated the source ephemerides and we discuss the orbital period evolution over a baseline of almost 17 years. The spectral analysis of these data will be reported elsewhere (Di Salvo et al. in prep).

\section[]{Observations and data reduction}
\subsection{XMM-Newton}
\label{sec:XMM}
We analysed a pointed \xmm{} observation of \saxj{} performed on 2015, April 11 (Obs.ID. 0724490201). During the observation the Epic-pn camera was operated in {\sc timing} mode (exposure time of $\sim105$ ks), while the RGS instrument was observing in spectroscopy mode (exposure time of $\sim80$ ks). Almost in the middle of the observation ($\sim40$ ks from the beginning) the satellite experienced a problem with the pointing system resulting in an $\sim$21 ks off-target observation (see Fig.~\cref{fig:lc_xmm}), that we excluded from the analysis. For this work we focused on the EPIC-pn (PN) data, that we extracted using the Science Analysis Software (SAS) v. 14.0.0 with the up-to-date calibration files, and adopting the standard reduction pipeline RDPHA \citep[see e.g.,][for more details on the method]{Pintore14}. We filtered the data in the energy range 0.3-10.0 keV, selecting single and double pixel events only (\textsc{pattern$\leq$4}). The average PN count rate during the observation was $\sim450$ cts/s, showing a clear decreasing trend from 550 cts/s at the beginning, down to 400 cts/s at the end of the observation (see Fig.~\ref{fig:lc_xmm}). We estimated a background mean count rate in the RAWX range [3:5] of the order of $\sim0.6$ cts/s in the energy range 0.3-10.0 keV. Moreover, we verified that the background region was not heavily contaminated by the source. No type-I burst episodes have been recorded in the PN data. 

\noindent
Fig.~\ref{fig:lc} shows the light curve of the 2015 outburst of the source monitored by \textit{Swift}-XRT (empty-circle points) and \textit{Swift}-BAT (black points). The green star represents the \xmm{} data taken roughly at the outburst peak. We corrected the PN photon arrival times for the motion of the Earth-spacecraft system with respect to the Solar System barycentre by using the \textsc{barycen} tool (DE-405 solar system ephemeris). We applied the best available optical position of the source \citep[][]{Hartman08} listed also in Tab.~\ref{tab:solution}.

\subsection{\nustar{}}
\begin{figure}
\centering
\subfloat[\xmm{} PN, background-subtracted light curve (0.3-10 keV) at 10-s time resolution of the 2015 outburst of \saxj{}. The drop-out in the middle of the observation is the result of $\sim21$ ks off-target pointing of the instrument.]{\label{fig:lc_xmm}\includegraphics[width=0.48\textwidth]{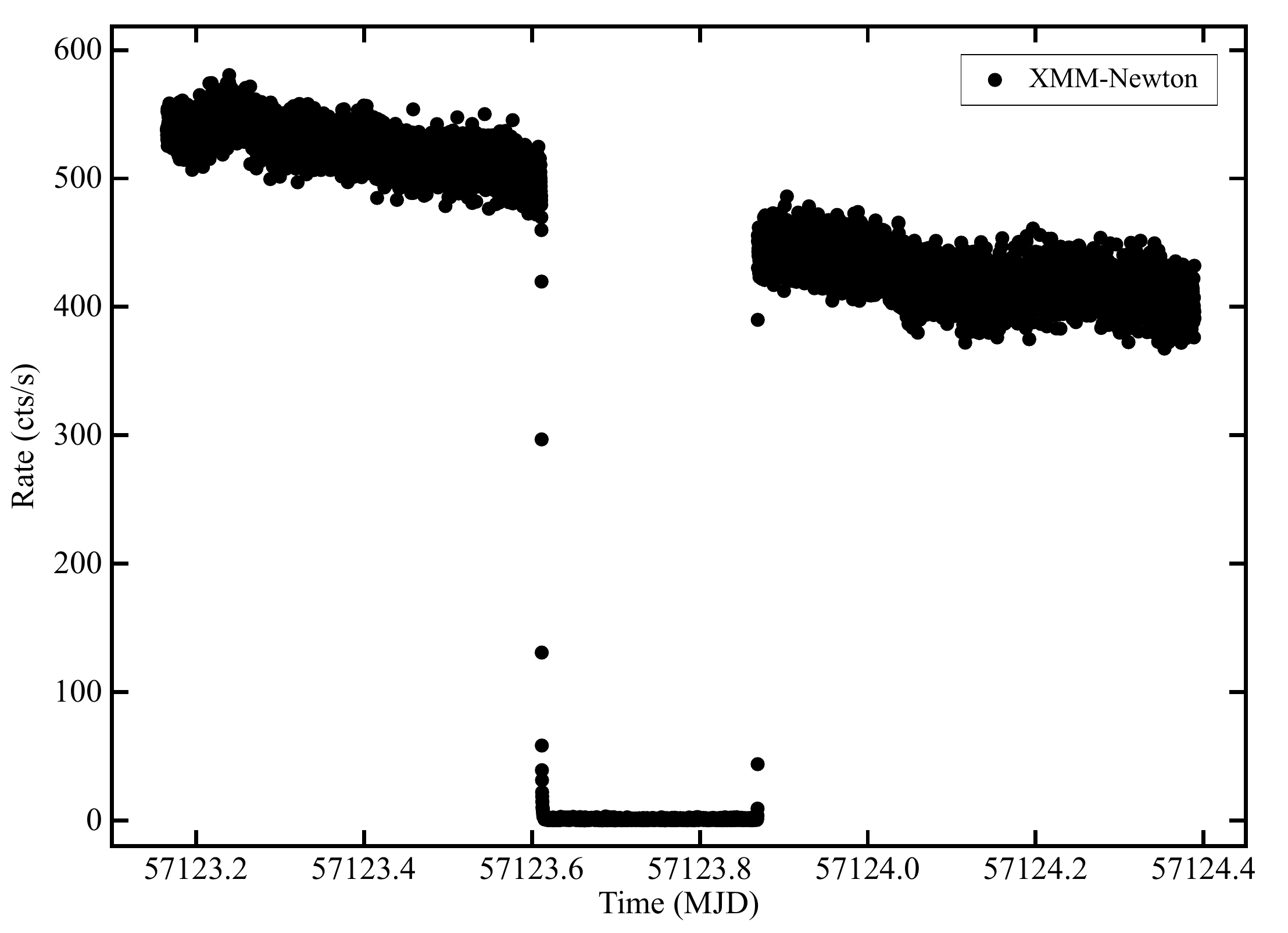} } 

\subfloat[\nustar{} FMPA, background-subtracted light curve (3$-$79 keV) at 10-s time resolution of the 2015 outburst of \saxj{}. After $\sim13$ ks from the beginning of the observation the source showed an X-ray burst.]{\label{fig:lc_nustar}\includegraphics[width=0.48\textwidth]{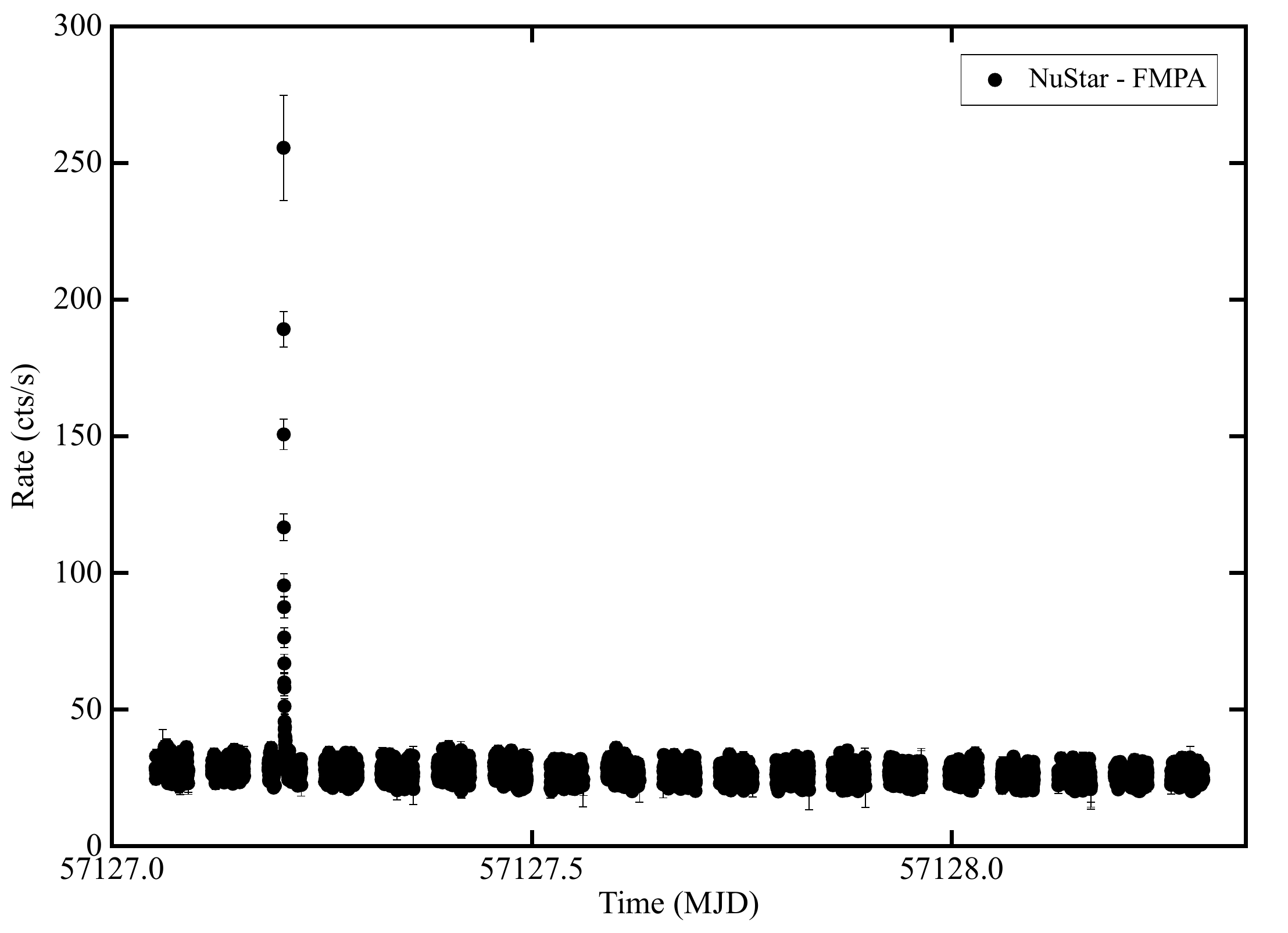} } 

\label{fig:lcurves}
\caption{Light curves of the observations analysed in this work.}
\end{figure}

\begin{figure}
\centering
\includegraphics[width=0.48\textwidth]{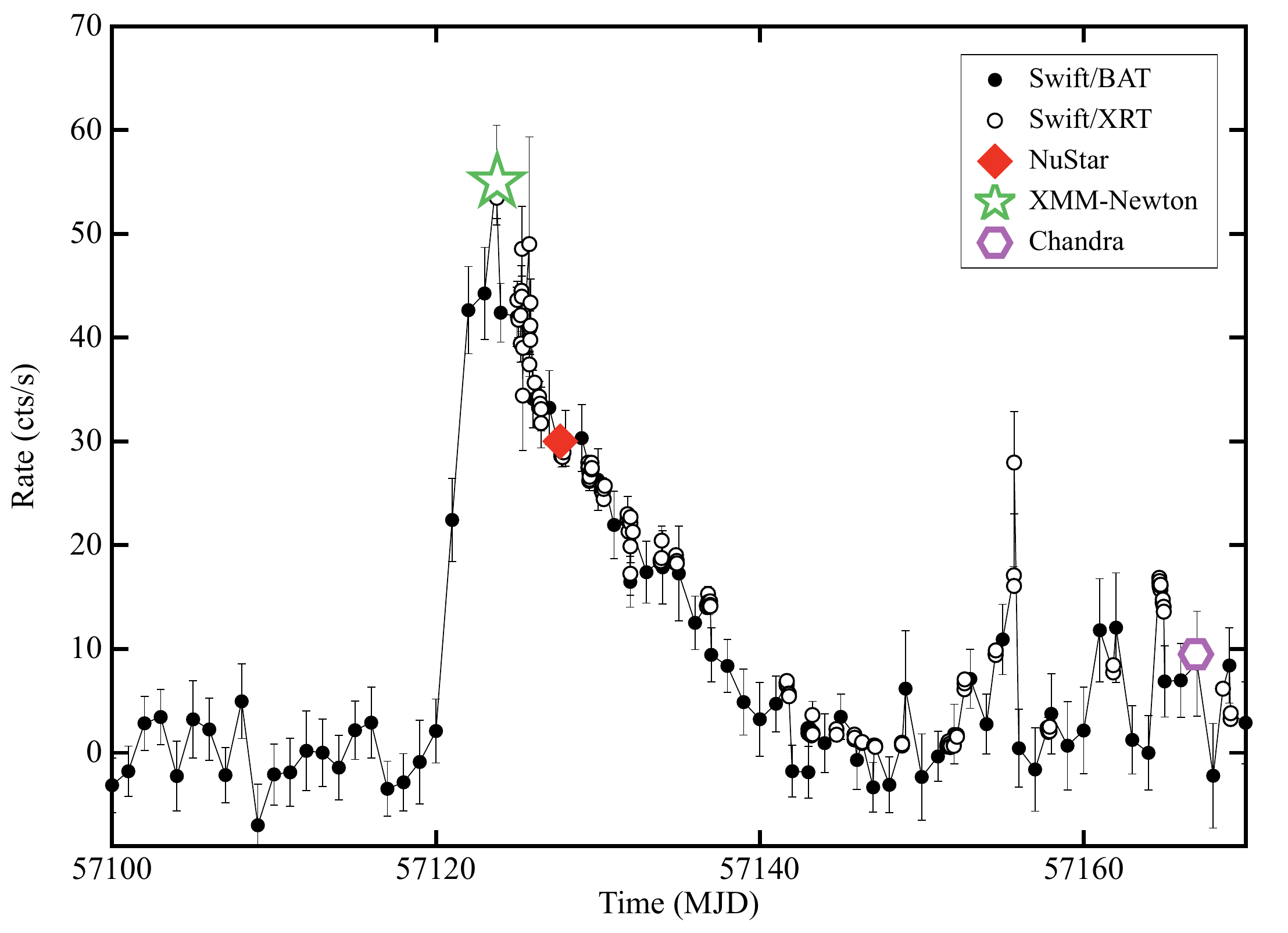}
\caption{Light-curve of the 2015 outburst of \saxj{} as observed by \textit{Swift}-XRT (open-circle points), and \textit{Swift}-BAT (black points). The green star and the red diamond represent the observations collected by \xmm{} and \nustar{}, respectively. The purple hexagon shows the \chandra{} observation reported by \citet{Patruno2016a}. Count rates from the instruments have been rescaled to match \textit{Swift}-XRT.}
\label{fig:lc}
\end{figure}

\saxj{} was observed by \nustar{} (Obs.ID. 90102003002) between 01:00 \texttt{UT} on 2015, April 15 and 07:30 \texttt{UT} on April 16. The red diamond in Fig.~\ref{fig:lc} shows the location of the \nustar{} observation with respect to the source outburst. We processed the events with the \nustar{} data analysis software (\textsc{nustardas}) version 1.5.1, resulting in an exposure time of $\sim49$ ks for each instrument. We filtered the source events from the FPMA and FPMB focal planes extracting a circular region of radius 100$''$ centered at the source position. The same extracting region, but centered far from the source, has been used to extract the background. Furthermore, combining the tool \textsc{nuproducts} and \textsc{lcmath} we extracted background-subtracted light curves for the two detectors (see Fig.~\ref{fig:lc_nustar}), characterised by an average count rate per instrument of $\sim28$ counts/s. During the observation a burst episode has been recorded. Solar System barycentre corrections were applied to the photon arrival times with the {\sc barycorr} tools (using DE-405 solar system ephemeris) as for \xmm{}.

\section{Data analysis and results}
\subsection{Timing analysis}
\label{sec:ta2015}
Starting from the timing solution reported by \citet[][]{Patruno12a} for the 2011 outburst of the source, we estimated the time delays $z(t)$ caused by the binary motion of the system under the hypothesis of almost circular orbits \citep[$e \ll 1$, see][for more details]{Burderi07}, through the formula:
\begin{eqnarray} 
\label{eq:bary}
\frac{z(t)}{c}= \frac{a \sin i}{c}\,\sin\Big(\frac{2\pi}{P_{orb}} \,(t-T_{\text{NOD}})\Big),
\end{eqnarray}
where $a \sin{\textit{i}/c}$ is the projected semi-major axis of the NS orbit in light seconds, $P_{orb}$ is the orbital period and $T_{\text{NOD}}$ is the time of passage at the ascending node. We extrapolated the starting value of the time of passage at the ascending node of the latest outburst ($T_{\text{NOD}, 2015}$) following the orbital evolution reported in their section 4.2 and assuming an orbital-period derivative of $\dot{P}_{orb, 2011}=3.5 \times 10^{-12}$ s/s \citep[see][and reference therein for more details.]{Patruno12a}
We then corrected the photon time of arrivals of the PN and \nustar{} events through the recursive formula  
\begin{eqnarray}
\label{eq:barygen} 
t + \frac{z(t)}{c} = t_{arr},
\end{eqnarray}
where $t$ is photon emission time, $t_{arr}$ is the photon arrival time to the Solar System barycentre. The correct emission times (up to an overall constant $D/c$ , where $D$ is the distance between the solar system barycenter and the barycenter of the binary system) are calculated by solving iteratively the aforementioned Eq.~(\ref{eq:barygen}), $t_{n+1} = t_{arr} - z(t_{n})/c$,
with $z(t)/c$ defined as in Eq.~(\ref{eq:bary}),
with the conditions $D/c = 0$, and $z(t_{n=0}) = 0$. We iterated until the difference between two consecutive steps (${\Delta t}_{n+1} = t_{n+1} - t_{n}$) is of the order of the absolute timing accuracy of the instrument used for the observations. In our case we set ${\Delta t}_{n+1}=1 \mu$s. 

Following the most updated long-term spin frequency evolution of \saxj{} \citep{Patruno12a}, we looked for pulsations performing epoch-folding search techniques of the whole observations using 16 phase bins, and starting with the spin frequency value $\nu_0$ = 400.97520998 Hz. We explored the frequency space around $\nu_0$ with steps of $10^{-8}$ Hz, for a total of 1001 steps. We found X-ray pulsation in both the PN and the \nustar{} observations at a mean frequency of $\nu=400.9752090(1)$ Hz and $\nu=400.975214(1)$ Hz, respectively. To estimate the error on the spin frequency we performed Monte Carlo simulations generating 100 data sets with the same properties of the real data such as, length, count rate, pulsation fractional amplitude and orbital modulation. Applying the method previously described we derived a mean frequency value for each simulated data set. We defined the $1\sigma$ uncertainty interval as the standard deviation of the spin frequency distribution from the simulation. 

\begin{figure*}
\centering
\includegraphics[width=0.9\textwidth]{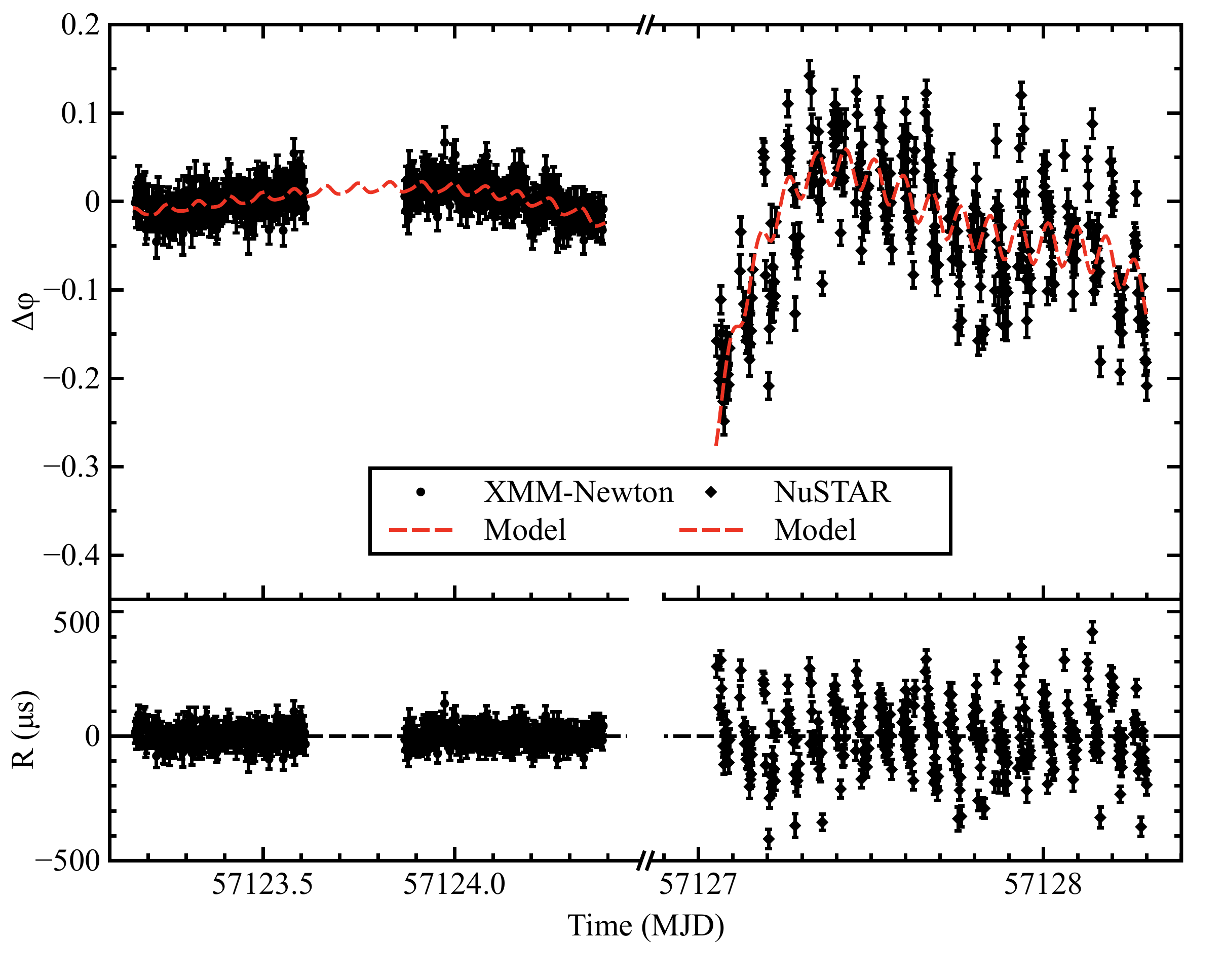}
\caption{\textit{Top panel -} Pulse phase delays as a function of time computed by epoch-folding the \xmm{} and \nustar{} observations at the spin frequency $\nu_0$ = 400.97520998 Hz, together with the best-fit model (red dotted line, see text). \textit{Bottom panel -} Residuals in $\mu s$ with respect to the best-fitting orbital solution.}
\label{fig:phase_fit}
\end{figure*} 

For the \nustar{} dataset, we repeated the analysis excluding the X-ray burst detected during the observation. We did not observe significant variation in terms of detectability of the pulse profile. Moreover, we investigated the presence of the coherent pulsation during the X-ray burst, finding evidence of X-ray pulsation with a statistical significance of $\sim4\sigma$ consistent within errors with the average spin frequency value of the \nustar{} observation reported in Tab.~\ref{tab:solution}. We decided not to exclude the X-ray burst from the timing analysis of the source. Although intriguing, a detailed analysis of the X-ray burst properties is beyond the scope of this work. 

We investigated, separately for \xmm{} and \nustar{}, the timing properties of the source by studying the evolution of the pulse phase delays computed on time intervals of approximately 200 seconds. We epoch-folded each segment in 16 phase bins at the spin frequency $\nu=400.9752091$ corresponding to the mean spin frequency during the PN observation with respect to the epoch $T_0=57123.1$ MJD. We modelled each pulse profile with a sinusoid of unitary period to determine the corresponding sinusoidal amplitude and the fractional part of phase residual. We selected only folded profiles with ratio between the sinusoidal amplitude and the corresponding $1\sigma$ error larger than 3. We detected pulsations on $\sim80\%$ and $\sim60\%$ of the intervals created from the PN and \nustar{} observations, respectively. We tried to fit the pulse profiles including a second harmonic component, but this component resulted statistically significant only in a small fraction of intervals ($\sim$ 6\% and $\sim$1\% of the intervals in PN and \nustar{}, respectively). The fractional amplitude of the signal varies between $\sim4\%$ and $\sim6\%$, with a mean value of $\sim4.5\%$ for the PN observation, while for \nustar{} it varies between $\sim5.5\%$ and $\sim10\%$, with a mean value of $\sim8\%$.

To determine a more accurate set of ephemeris we performed timing analysis of the combined set of PN and \nustar{} observations by fitting the time evolution of the pulse phase delays with the following models: 
\begin{eqnarray}
\label{eq:ph}
\begin{cases}
\Delta \phi_{PN}(t)=\sum\limits_{n=0}^{3} \frac{C_n}{n!}(t-T_0)^n+R_{orb}(t),\\
\Delta \phi_{NS}(t)=\sum\limits_{n=0}^{7} \frac{D_n}{n!}(t-T_0)^n+R_{orb}(t),
\end{cases}
\end{eqnarray}
where the first element of both equations represent a polynomial function used to model phase variations superposed to the residual orbital modulation $R_{orb}(t)$ caused by differences between the \emph{real} set of orbital parameters and those used to correct the photon time of arrivals \citep[see e.g.,][]{Deeter81}. The order of the polynomial function varied between the datasets depending on the fluctuations of the pulse phases. We note that no phase-lock timing analysis of the \xmm{} and \nustar{} datasets has been possible given the low accuracy of the timing solutions from the single observations combined with the temporal gap between them ($\sim$3 days). Furthermore, the presence of a time drift on the internal clock of the \nustar{} instrument \citep{Madsen15}, which affects the observed coherent signal making the spin frequency value significantly different compared to the PN, represents a strong limitation on the phase-connected variability timing techniques.  
We then emphasise that the two models in Eq.~\ref{eq:ph} only share the same orbital parameters component $R_{orb}(t)$, meaning that the orbital parameters will be linked during the fit of the two datasets. This method has the advantage (with respect to model each observation separately) to improve the accuracy of parameters such as the orbital period and the time of passage to ascending node. If a new set of orbital parameters is found, photon time of arrivals are corrected using Eq.~\ref{eq:bary} and pulse phase delays are created and modelled with Eq.~\ref{eq:ph}. We repeated the process until no significant differential corrections were found for the parameters of the model. We reported the best-fit parameters in Tab.~\ref{tab:solution}, while in Fig.~\ref{fig:phase_fit} we showed the pulse phase delays of the two instruments with the best-fitting models (top panel), and the residuals with respect to the models (bottom panel). The value of $\tilde{\chi}^2\sim1.62$ (with 696 degrees of freedom) shows that the model well fits the pulse phase delays. However, the large distribution of the residuals from the \nustar{} observation (bottom-right panel) clearly shows the presence of a peculiar timing noise in these data, likely ascribable to the instrumental issue mentioned before \citep{Madsen15}.

\begin{figure}
\centering

\subfloat[\textit{XMM-Newton $-$} Pulse phase delays as a function of time for the fundamental (black-filled dots) and the second harmonic (black-empty dots) components of the source spin frequency. The solid and dotted green lines represent the best-fitting models for the fundamental and the second harmonic, respectively. ]{
        \label{fig:fit_xmm}
        \includegraphics[width=0.48\textwidth]{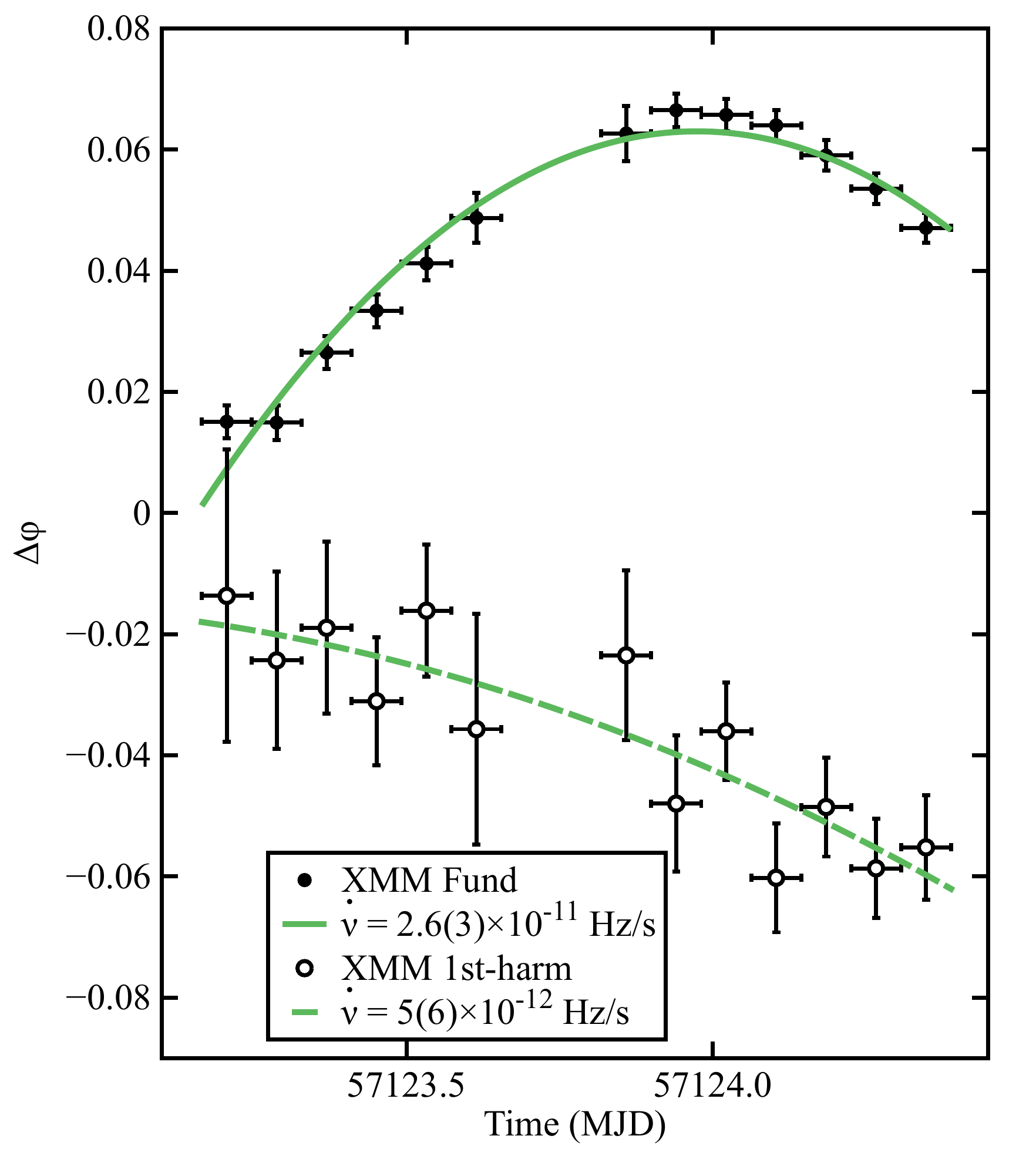} } 

\subfloat[\textit{NuStar$-$} Pulse phase delays as a function of time for the fundamental (red-filled stars) and the second harmonic (red-empty stars) components of the source spin frequency. The solid and dotted green lines represent the best-fitting models for the fundamental and the second harmonic, respectively.]{
        \label{fig:fit_nustar}
        \includegraphics[width=0.48\textwidth]{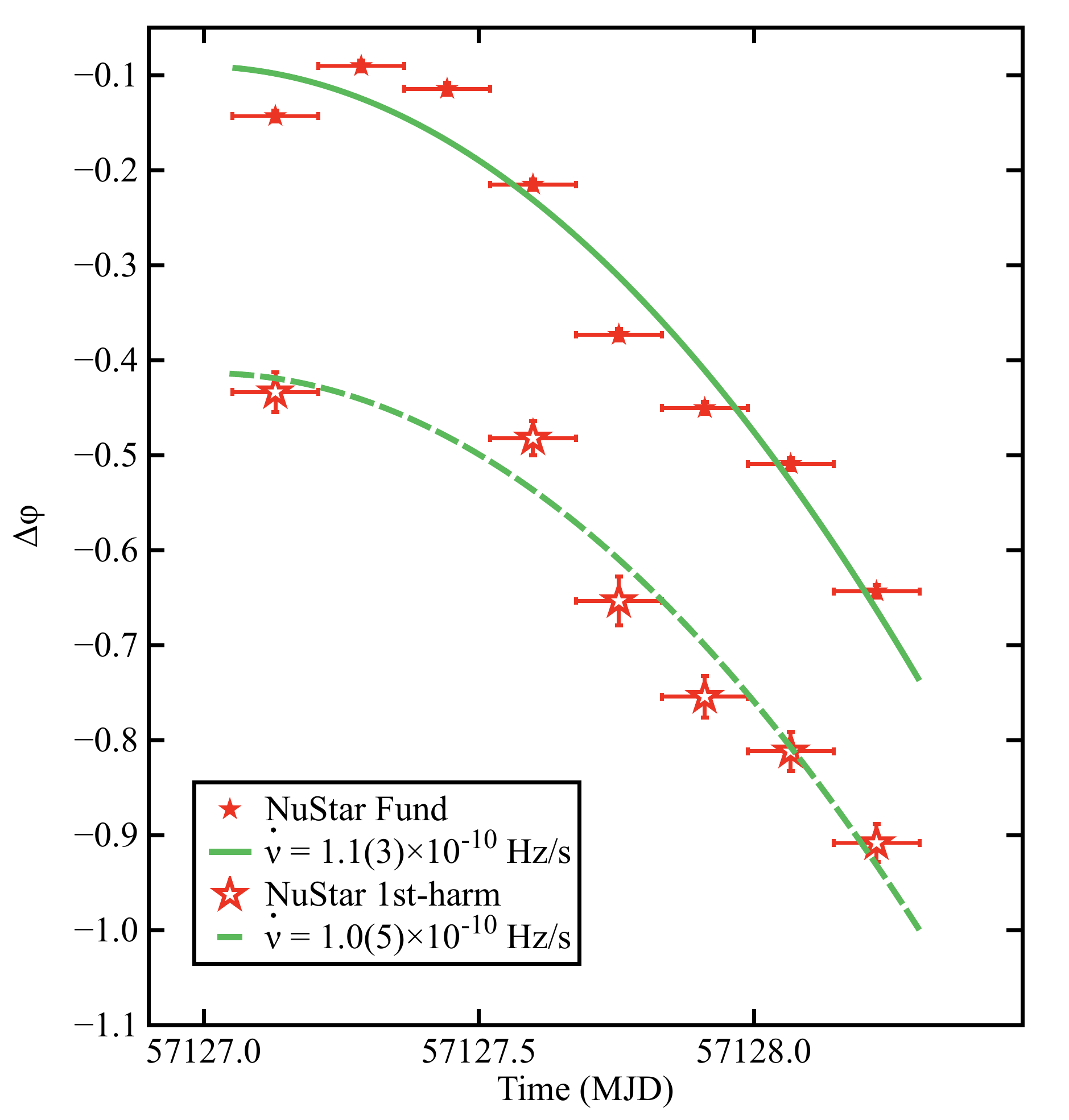} } 
        
\caption{Time evolution of the pulse phase delays for the two datasets and corresponding best-fitting models.}
\label{fig:profiles}

\end{figure}

With the updated set of ephemerides reported in Tab.~\ref{tab:solution}, we corrected the events of the PN and \nustar{} observations. We then epoch-folded, for each observation, long intervals of data (between one and two orbital periods, depending on the statistics of the data) in order to have pulse profiles with significant fundamental and second harmonic components. To investigate the evolution of the spin frequency we fitted the pulse phase delays as a function of time with the expression:
\begin{eqnarray}
\label{eq:ph_fit}
\Delta \phi(t)=\phi_0+\Delta \nu\,(t-T_0)+\frac{1}{2}\dot{\nu}\,(t-T_0)^2,
\end{eqnarray}  
where $\phi_0$ is a constant phase, $\Delta \nu=(\nu_0-\nu)$ represents the difference between the frequency at the reference epoch and the spin frequency used to epoch-fold the data, $\dot{\nu}$ is spin frequency derivative, and $T_0$ represents the reference epoch for the timing solution. For each observation, we modelled with Eq.~\ref{eq:ph_fit} the phase delays obtained from both the fundamental and the second harmonic components. We reported the best-fit parameters in Tab.~\ref{tab:solution}, while in Fig.~\ref{fig:fit_xmm}$-$\ref{fig:fit_nustar} we showed the pulse phase delays with the best-fitting models for PN, and \nustar{}, respectively. In Tab.~\ref{tab:solution} we also reported the mean spin frequency values obtained by modelling the phase delays of the fundamental component with a constant model.
We measured significant spin frequency derivative values from the phase delays of the fundamental component of both the dataset with values of $2.6(3)\times 10^{-11}$ Hz/s and $1.1(3)\times 10^{-10}$ Hz/s for PN and \nustar{}, respectively. Moreover, it is worth noting that the PN dataset (Fig.~\ref{fig:fit_xmm}) shows a clear mismatch between the time evolution of the fundamental and second harmonic components, while in \nustar{} (Fig.~\ref{fig:fit_nustar}) the two components show an overall behaviour consistent within errors. 
\begin{table*}

\begin{tabular}{l | c  c  }
Parameters             & \xmm{} & \nustar{}  \\
\hline
\hline
R.A. (J2000) &  \multicolumn{2}{c}{$18^h08^m27^s.62$}\\
DEC (J2000) & \multicolumn{2}{c}{$-36^\circ58'43''.3$}\\
Orbital period $P_{orb}$ (s) & \multicolumn{2}{c}{7249.143(4)}\\
Projected semi-major axis a sin\textit{i/c} (lt-ms) &\multicolumn{2}{c}{62.810(3)} \\
Ascending node passage $T_{\text{NOD}}$ (MJD) & \multicolumn{2}{c}{57123.060218(3)}\\
Eccentricity (e) & \multicolumn{2}{c}{< 4$\times 10^{-4}$}\\
$\chi^2$/d.o.f. &\multicolumn{2}{c}{1126.2/696}\\
\hline
\hline
Fundamental\\
\hline
\hline
Spin frequency $\nu_0$ (Hz) &400.9752089(1)& 400.975214(1)\\
$\chi^2$/d.o.f. &{184.4/11} &{706.9/6} \\
\hline
Spin frequency $\nu_0$ (Hz) &400.9752075(2)& 400.975208(2)\\
Spin frequency 1st derivative $\dot{\nu}_0$ (Hz/s) &2.6(3)$\times 10^{-11}$&1.1(3)$\times 10^{-10}$\\
$\chi^2$/d.o.f. &{20.1/10} &{333.8/5} \\
\hline
2nd Harmonic\\
\hline
\hline
Spin frequency $\nu_0$ (Hz) &400.9752098(1)& 400.975213(1)\\
$\chi^2$/d.o.f. &{5.5/11} &{46.3/5} \\
\hline
Spin frequency $\nu_0$ (Hz) &400.9752094(4)& 400.975208(3)\\
Spin frequency 1st derivative $\dot{\nu}_0$ (Hz/s) &7(7)$\times 10^{-12}$&0.9(5)$\times 10^{-10}$\\
$\chi^2$/d.o.f. &{5.1/10} &{19.9/3} \\
\hline
\end{tabular}
\caption{Orbital parameters of \saxj{} combining the timing analysis of the \xmm{} and \nustar{} observations from the 2015 outburst of the source. The reference epoch for the solution is T$_0$=57123.1 MJD. Spin frequency and spin frequency derivative estimates are obtained by fitting independently the two observations. Errors are at 1$\sigma$ confidence level. The reported X-ray position of the source has a pointing uncertainty of 0.15$''$ \citep[][]{Hartman08}.}
\label{tab:solution}
\end{table*}

Taking into account the uncertainties on the source position, we estimated the systematic uncertainties induced on the spin frequency correction $\delta \nu$, and the spin frequency derivative $\dot{\nu}$. Using the expression of the residuals induced by the motion of the Earth for small uncertainties on the source position $\delta_{\lambda}$ and $\delta_{\beta}$ expressed in ecliptic coordinates $\lambda$ and $\beta$ \citep[see, e.g.,][]{Lyne90}, we derived the expressions: $\sigma_{\nu_{pos}}\leq \nu_0\,y\,\sigma_{\gamma}(1+\sin^2\beta)^{1/2}2\pi/P_{\oplus}$, and $\sigma_{\dot{\nu}_{pos}}\leq \nu_0\,y\,\sigma_{\gamma}(1+\sin^2\beta)^{1/2}(2\pi/P_{\oplus})^2$, where $y=r_E/c$ is the semi-major axis of the orbit of the Earth in light-seconds, $P_{\oplus}$ is the Earth orbital period, and $\sigma_{\gamma}$ is the positional error circle. Considering the positional uncertainty of $0.15''$ reported by \citet{Hartman08}, we estimated $\sigma_{\nu_{pos}} \leq 1\times 10^{-10}$ Hz, and $\sigma_{\dot{\nu}_{pos}} \leq 2\times 10^{-17}$ Hz/s, respectively. We added in quadrature these systematic uncertainties to the statistical errors of $\nu_0$, and $\dot{\nu}$ estimated from the timing analysis.

\begin{figure}
\centering
\includegraphics[width=0.48\textwidth]{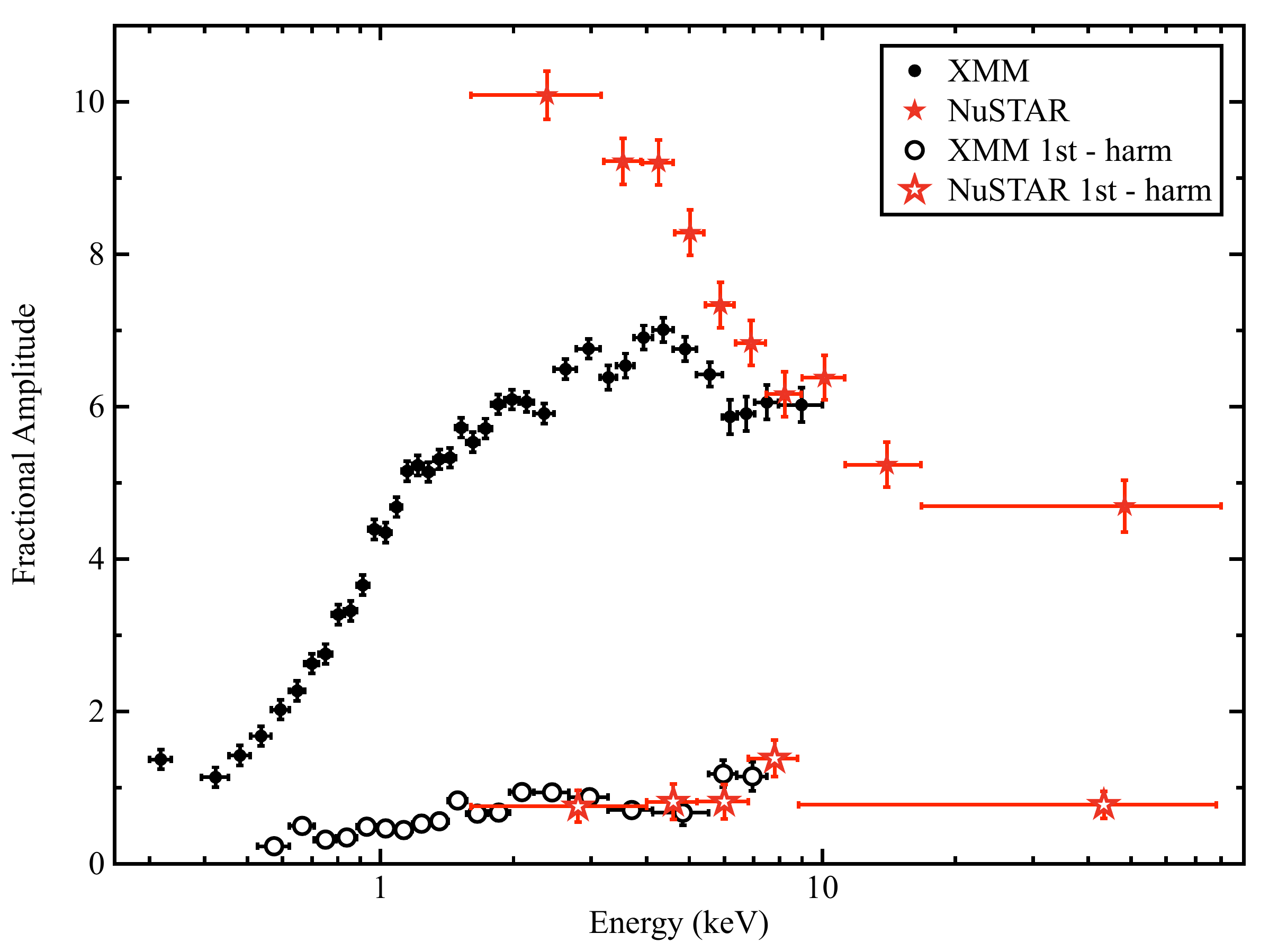}
\caption{Evolution of the pulse profile fractional amplitude of the fundamental and first overtone used to model the pulse profile obtained from the \xmm{} (black-filled dots and black empty dots) and the \nustar{} (red-filled stars and red empty stars) observations as a function of energy.}
\label{fig:amp_vs_energy}
\end{figure}

Finally we investigated the properties of the pulse profile as a function of energy, dividing the PN energy range between 0.3 keV to 10 keV into 38 intervals, and the \nustar{} energy range between 1.6 keV and 80 keV in 10 intervals. We adjusted the width of the energy bins in order to be able to significantly detect the pulsation. We modelled the background-subtracted pulse profile with two sinusoidal components (fundamental and second harmonic) for which we calculated the fractional amplitudes for each energy selection. Fig.~\ref{fig:amp_vs_energy} shows the dependence of the fractional amplitude of the pulse profile as a function of energy. The PN fractional amplitude of the fundamental component increases from $\sim1\%$ at around 0.4 keV up to $\sim7\%$ at 5 keV, after that it starts decreasing reaching the value of $\sim6\%$ at 10 keV. It is interesting to note that, in addition to the decreasing trend, the fractional amplitude seems to drop in correspondence to the Iron line energy range (6.4$-$7 keV) \citep[a similar behaviour has been observed during the 2008 outburst of the source, see e.g., ][]{Patruno09b}. The \nustar{} fractional amplitude of the fundamental component basically shows a decreasing trend between $\sim10\%$ at around 2 keV and $\sim5\%$ up to 80 keV. Furthermore, in Fig.~\ref{fig:amp_vs_energy} can be clearly seen the discrepancy between the fundamental components (black-filled points and red-filled stars) of the two instruments, both in terms of maximum fractional amplitude detected and overall amplitude-energy trend observed. Differently from the fundamental, the PN second harmonic shows an increasing trend from $\sim0.2\%$ at 0.5 keV up to $\sim1\%$ at 10 keV. In order to significantly detect the \nustar{} second harmonic we created new energy selections resulting on 5 amplitude measurements reported with red-empty stars in Fig.~\ref{fig:amp_vs_energy}. The fractional amplitude trend is more or less constant as a function of energy around the mean value $0.7\%$.

\begin{figure}
\centering
\includegraphics[width=0.48\textwidth]{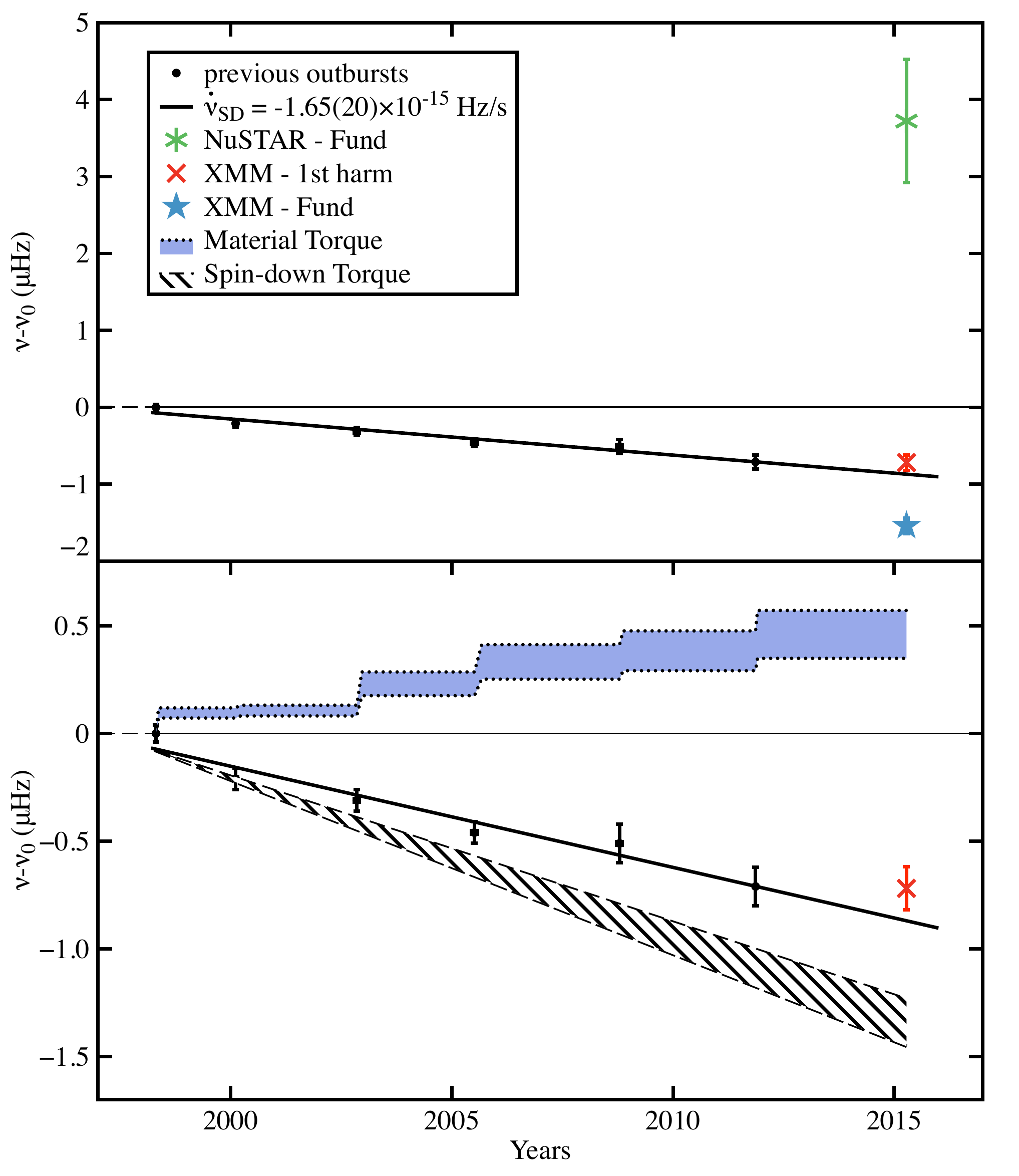}
\caption{{\it Top panel:} Secular evolution of the spin frequency of \saxj{} for the 1998--2015 baseline. Black points represent the frequency measurements of the previous outbursts reported by \citet{Patruno12a}. Green and blue points represent, respectively, the \nustar{} and \xmm{} frequency measurements during the 2015 outburst relative to the fundamental component of the X-ray pulsation. The red point represents the frequency measurement relative to the second harmonic component of the X-ray pulsation observed by \xmm{}. Frequencies are rescaled relative to value from the 1998 outburst ($\nu_0 = 400.97521052$ Hz). Solid line represents the best-fitting model of the spin frequency values observed in the outbursts between 1998 and 2011 reported by \citet{Patruno12a}. Errors are reported at 1$\sigma$ confidence level. {\it Bottom panel:} Schematic representation of the spin-up and spin-down effects acting on the NS spin frequency caused by the material torque and the rotating-magnetic dipolar torque during the outbursts of \saxj{}. Blue shaded region represents the spin-up effect caused by the material torque calculated assuming magnetic channeling in the accretion disc truncated at the NS radius and at the co-rotation radius. The hatched region represents the hypothetical magnetic dipolar spin-down effect required to describe the measured spin-down effect (black solid line) after taking into account the spin-up from the material torque.}
\label{fig:spin_evo}
\end{figure}

\begin{figure*}
\centering
\includegraphics[width=0.8\textwidth]{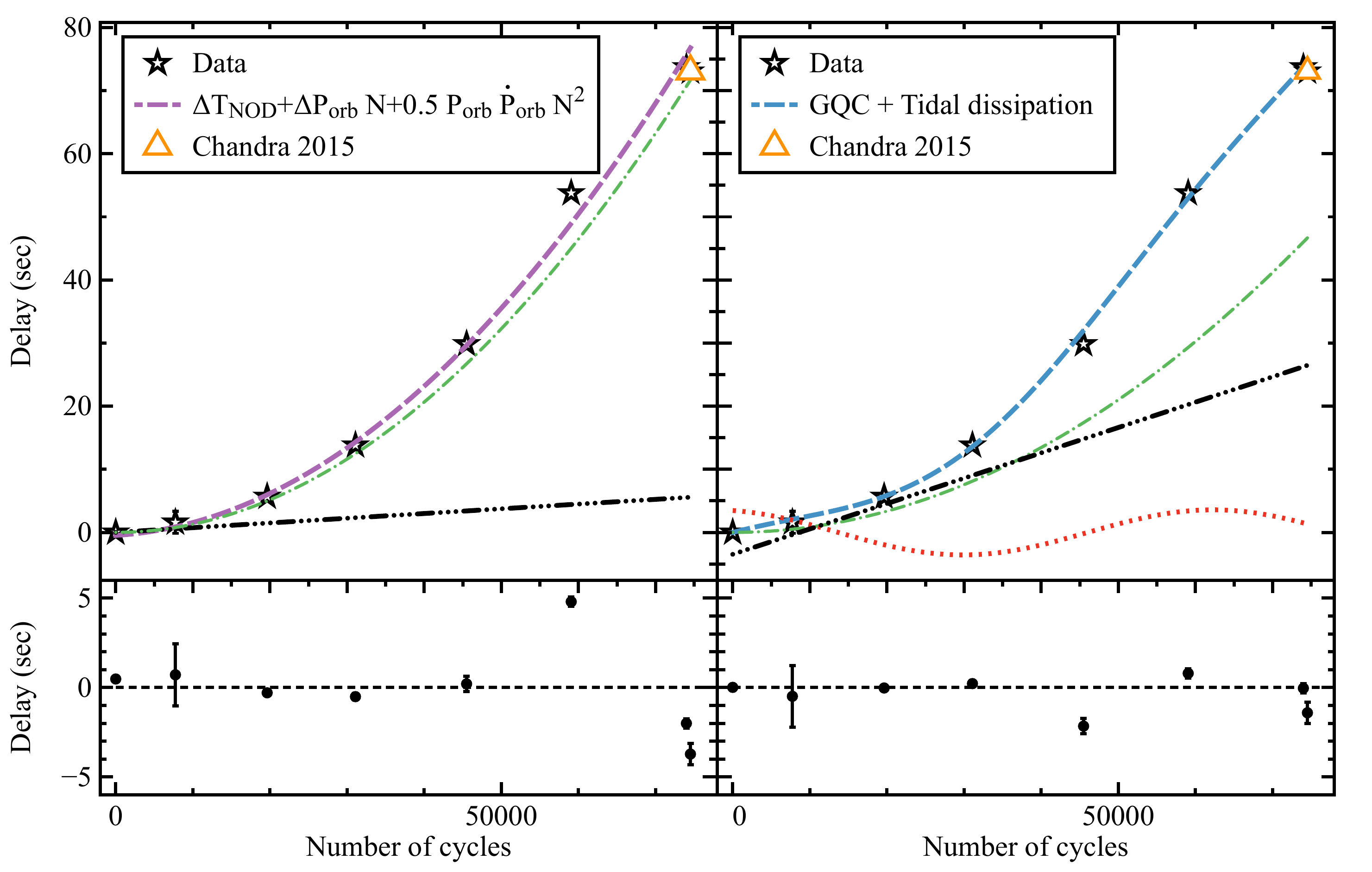}
\caption{\textit{Left-top panel -} Differential correction $\Delta T_{\text{NOD}}$ to the time of passage at the ascending node for the seven outbursts shown by \saxj{}. Each point is computed with respect to the beginning of the first outburst occurred in 1998 (see Sec.~\ref{sec:orb} for more details). The dashed purple line represents the best-fitting quadratic model used to fit the data. Dot-dot-dot-dashed black line and dot-dashed green line represent the linear and the quadratic components of the model, respectively. \textit{Left-bottom panel -} Residuals with respect to the best-fitting quadratic model. \textit{Right-top panel -} Same differential correction $\Delta T_{\text{NOD}}$ shown in the left panel, but fitted with the gravitational quadrupole coupling (GQC) model powered by a tidal dissipation mechanism (blue). Dot-dot-dot-dashed black line, dot-dashed green line and dotted red line represent the linear, the quadratic and the sinusoidal components of the model, respectively. \textit{Right-bottom panel -} Residuals with respect to the a gravitational quadrupole coupling (GQC) model powered by a tidal dissipation mechanism. For comparison we reported the value $T_{\text{NOD}}$ from the 2015 outburst reported by \citet{Patruno2016a}.}
\label{fig:tstars}
\end{figure*}

\section{Discussion}

We have presented an updated timing solution for the accreting millisecond X-ray pulsar \saxj{} obtained from the timing analysis of the \xmm{} and \nustar{} observations of its 2015 outburst. The new set of orbital parameters is compatible within the errors with the timing solution obtained from the analysis of the \chandra{} observation of the same outburst reported by \citet{Patruno2016a}.

\subsection{Secular spin evolution}
\label{sec:spin}
As mentioned in the previous sections, the source has been observed in outburst seven times since its discovery \citep[see e.g. ][]{Hartman08, Burderi09, Hartman09b, Patruno12a, Patruno2016a}. To evaluate its secular spin frequency variation we started by considering the six constant-pulse frequencies estimated from the outbursts between 1998 and 2011 \citep[reported by][see their Fig.~2, bottom panel]{Patruno12a} and shown in the top panel of Fig.~\ref{fig:spin_evo} (black points). In addition we plotted the spin frequency measurements obtained from the timing analysis of the pulse phases of the fundamental component from the 2015 outburst obtained with \xmm{} (blue point) and \nustar{} (green point). Both values significantly deviate from the expected trend drawn from the secular evolution of the previous outbursts (black solid line). The frequency discrepancy shown by \nustar{} most likely represents an artefact that originates from the time drift of the internal clock of the instrument \citep[see e.g., ][]{Madsen15}. The presence of a spurious spin frequency derivative discussed in the previous section and reported in Fig.~\ref{fig:fit_nustar}, can significantly change the spin frequency measurements explaining the large discrepancy with respect to both the quasi-simultaneous spin frequency obtained with \xmm{} and the secular evolution predicted from the previous outbursts. A large discrepancy is also observed for the spin frequency value obtained from the fit of the pulse phase fundamental component (ignoring spin frequency derivatives) of the \xmm{} observation. However, we note that, as reported in Fig.~\ref{fig:fit_xmm}, a large spin-up is detected from the analysis of this component within the $\sim105$ ks of exposure. This value is almost two order of magnitude larger with respect to the value reported by \citet{Burderi06} for the 2002 outburst of the source, as well as to the upper limits reported for the other outbursts \citep[see e.g.,][]{Hartman08, Hartman09b, Patruno12a}. Interestingly, we note that the frequency estimated from the timing analysis of the second harmonic component (red point) is significantly inconsistent with that estimated from the fundamental component and it falls very close (less than 2$\sigma$) from the predicted secular evolution. Moreover, contrary to the fundamental component, no significant spin frequency derivative is detectable (with a $3\sigma$ upper limit of $2.3\times 10^{-11}$ Hz s$^{-1}$, see Fig.~\ref{fig:fit_xmm}). The different time evolution between the fundamental and the first overtone of the pulse phase delays of \saxj{} has been already reported and extensively discussed for the previous outbursts of the source \citep[see e.g., ][]{Burderi06, Hartman08, Patruno09c}, however, no general agreement has been reached regarding the mechanism behind this phenomenon. We suggest that both the frequency shift and the unusually large spin frequency derivative observed from the analysis of the pulse phase fundamental component are connected to the known phase-jumping phenomenon, which most likely does not represent a real evolution of the NS spin \citep[see e.g., ][]{Patruno09c, Riggio2011a}. Although very intriguing, this subject is quite complex and requires extensive discussions that fall outside the focus of this work and they will be properly investigated in a follow up project. 

To investigate the secular spin evolution of \saxj{}, we then consider as a good proxy of the spin frequency at the beginning of the 2015 outburst, the value estimated from the analysis of the pulse phase delays second harmonic component. We modelled the seven spin frequency values with a linear function. From the fit we obtain a $\chi^2=8.1$ with 5 d.o.f. and a spin frequency derivative of $\dot{\nu}=-1.5(2)\times 10^{-15}$ Hz/s, consistent with the value reported by \citet{Patruno12a}.   
Equating the rotational-energy loss rate to the rotating magnetic dipole emission, we can infer the strength of the magnetic field at the NS polar caps. Following \citet{Spitkovsky2006a} and assuming a rotating dipole in the presence of matter, we can express the NS magnetic dipole moment as
\begin{equation}
\mu \simeq 1.03\times10^{26}\left(\frac{1}{1+\sin^2{\alpha}}\right)^{-1/2} I_{45}^{1/2}\nu_{401}^{-3/2}\dot{\nu}_{-15}^{1/2}\,\,\,\, \textrm{G\,cm$^3$}
\label{eq:mag}
\end{equation}
where $\alpha$ is the angle between the rotation and magnetic axes, $I_{45}$ is the moment of inertia of the NS in units of $10^{45}$ g cm$^2$,  $\nu_{401}$ is the NS spin frequency in units of 401 Hz, $\dot{\nu}_{-15}$ is the spin-down frequency derivative in units of $10^{-15}$ Hz/s. As suggested by \citet{Poutanen03}, the shape of the pulse profiles investigated during the 1998 outburst seem to favour a small misalignment between the magnetic hot spot and the rotational pole, with the angle $\alpha$ ranging between 5 and 20 degrees. Combining the constraints on the magnetic field geometry with our estimates of the spin frequency and its secular spin-down derivative, we can constrain the NS magnetic moment in the range $1.4\times10^{26}\,\,\, \textrm{G\,cm$^3$}< \mu < 1.5\times10^{26}\,\,\, \textrm{G\,cm$^3$}$.
Adopting the FPS equation of state \citep[see e.g.,][]{Friedman1981a, Pandharipande1989a} for a 1.4 M$_\odot{}$ NS, we estimate a NS radius of $R_{NS}=1.14\times10^{6}$ cm. Defining the magnetic field strength at the magnetic caps as $B_{PC}= 2 \mu/R_{NS}^{3}$, we obtain the magnetic field ranging between $1.9\times 10^{8}\,\,\, \textrm{G} <B_{PC}<2\times 10^{8}\,\,\, \textrm{G}$, which is consistent with the value reported by \citet{Patruno12a} as well as consistent with the value estimated from the detection of a relativistically broadened iron line during the 2008 outburst \citep[$B_{PC}=2-4 \times 10^{8}$ G, see e.g.,][]{Cackett09, Papitto09}. On the other hand, this value is almost a factor of two smaller compared with the magnetic field estimated by \citet{Burderi06} from the spin-down torque measured at the final stages of the 2002 outburst of the source.

It is worth noting that, even though \saxj{} accreted matter during the latest seven monitored outbursts, the overall secular spin evolution clearly show a modest slowdown of the NS. Assuming standard accretion torque theory \citep[see e.g.,][]{Pringle1972a}, starting from the amount of matter accreted onto the NS surface during an outburst, we can estimate the spin variation during the outbursts. We can define the spin frequency variation due to the accreting torque as
\begin{equation}
\Delta \nu = \frac{\Delta M \sqrt{G m_1 R_{\textrm{IN}}}}{2 \pi I}
\end{equation}
where $\Delta M$ represents the amount of matter accreted onto the NS, m$_1$ and $I$ represent the mass and the moment of inertia of the NS, respectively and $R_{\textrm{IN}}$ is the disc radius at which the matter is channelled by the magnetic field lines and forced to accrete onto the NS magnetic poles. A rough estimate of the spin-up variation can be obtained considering $\Delta M\simeq \Delta E_{\textrm{bol}}(G m_1/R)^{-1}$, where $\Delta E_{\textrm{bol}}$ represents the amount of energy released by \saxj{} during each outbursts. To quantify the spin frequency shift occurred between the beginning of the 1998 and the 2015 outbursts (same interval used to quantify the spin-down derivative), we consider the amount of energy released in the six outbursts happened in between. Combining the X-ray light curves reported in literature \citep[see e.g.,][]{Hartman08, Hartman09b, Patruno12a} and assuming that no outburst has been missed, we can estimate the energy released during the outburst activity of \saxj{} between 1998 and 2015 as $\Delta E_{\textrm{bol}}\simeq3.2\times10^{43}$ erg. The latter value has been inferred assuming a source distance of $d=3.5$ kpc \citep{Galloway2006a} and a bolometric correction factor of 2.12 \citep[see e.g.,][and references therein]{Hartman08} to derive the bolometric luminosity from the observed X-ray luminosity. Moreover, we considered as possible extreme values of the truncation disc radius during the outburst phase, the NS radius $R_{NS}=1.14\times10^{6}$ cm and the co-rotation radius $R_{\text{CO}}=3.08\times10^{6}$ cm (radius at which the Keplerian frequency equals the NS spin frequency). We then estimated that the frequency spin-up caused by the material torque of the 1998-2011 outbursts ranges between $\sim0.35 \mu$Hz and $\sim0.56 \mu$Hz. In Fig.~\ref{fig:spin_evo} (bottom panel), the blue shaded region represents the time evolution of the spin frequency as a consequence of the angular momentum accreted during the outburst phases of the source. If the material torque described above is correct, we should re-evaluate the spin-down torque acting onto the NS. Assuming the magnetic dipole radiation emission during the X-ray quiescence as the main driver of the NS spin-down, the spin-down derivative can be estimated combining the observed secular frequency shift (almost $-0.8\mu$Hz over 17 years of observations) with that caused by the material torque (ranging between $0.35 \mu$Hz and $0.56 \mu$Hz, considering the NS radius and the co-rotation radius as possible truncation radii of the disc). The spin-down derivative required to compensate the estimated spin-up material torque and to produce the observed spin-down effect should range between $-2.2\times 10^{-15}$ Hz/s and $-2.7\times 10^{-15}$ Hz/s, a factor of almost 1.5 and 1.8 higher, respectively, than the spin-down frequency derivative estimated ignoring the material torque. Using Eq.~\ref{eq:mag} and adopting the FPS equation of state previously described, we can estimate the NS dipole magnetic moment, hence the magnetic field at the magnetic caps in the range $2.2\times 10^8 \textrm{G} <B_{PC}< 3\times 10^8 \textrm{G}$, very similar to the value suggested by \citet{Burderi06}. However, we note that other mechanisms such as magnetic propeller torque and gravitational radiation torque \citep[see e.g.,][and references therein for more details]{Hartman08} could contribute to the spin-down torque required to describe the observed secular spin evolution of \saxj{}.

\subsection{Pulse energy dependence}

\saxj{} shows an interesting behaviour of its pulse profile as a function of energy. As shown in Fig.~\ref{fig:amp_vs_energy}, the pulse fractional amplitude estimated from the fundamental component varies significantly between the \xmm{} and the \nustar{} observations of the source. We can exclude instrumental effects between the two satellites to explain such a discrepancy. Indeed, phase-coherence analyses on almost simultaneous \xmm{} and \nustar{} have been carried out on other sources \citep[see e.g.][for the AMXP \igrj{}]{Sanna2017b} and they did not show any peculiar difference between the two detectors. Therefore, a possible explanation may be a significant source state variation occurred in the three days gap between the two observations. A detailed study of this phenomenon is beyond the scope of this work, however, further investigations on the subject will be reported in a follow up paper focusing on the spectral properties of \saxj{} during its 2015 outburst (Di Salvo et al. in prep.).

The peculiar trend showed by the fundamental component is consistent with that observed by the \rxte{} during the 1998 outburst while it significantly differs from that of the 2002 \citep[see e.g.,][]{Cui98b, Falanga07b, Hartman09a}. A similar behaviour has been observed for the other AMXPs such as \igrj{} \citep{Falanga05b, Sanna2017b}, \igj{} \citep{Papitto10} and \xt{} \citep{Falanga07b}, although for the latter a good coverage of the trend below 4 keV is lacking. On the other hand, the constant energy dependence of the fractional amplitude inferred from the second harmonic component seems to remain unvaried over the outbursts \citep[see e.g.,][]{Cui98b, Hartman09a}. The mechanism responsible for the complex energy spectrum observed for the AMXPs fractional amplitude is still under discussion, however processes such as strong comptonisation of the beamed radiation have been proposed to explain the hard spectrum of the pulsation observed in several systems \citep[see e.g,][]{Falanga07b}. Alternatively, \citet{Muno02, Muno03}, proposed the presence of a hot-spot region emitting as a blackbody with a temperature significantly different with respect to the neutron star surface as the mechanism responsible for the increasing pulse amplitude with energy in the observer rest frame, compatible to what observed for systems such as Aql X-1 \citep{Casella08}, SWIFT J1756.9$-$2508 \citep{Patruno10b}, XTE J1807$-$294 \citep{Kirsch04} and SAX J1748.9$-$2021 \citep{Patruno09a, Sanna2016a}.

\subsection{Orbital period evolution} 
\label{sec:orb}

During the last seventeen years \saxj{} has been detected in outburst seven times, and for each of these phases an accurate set of ephemerides has been produced. In order to investigate the evolution of the orbital period of the source we study the correction on the NS passage from the ascending node $\Delta T_{\text{NOD}}$ (with respect to the beginning of the 1998 outburst) for each outburst, as a function of the orbital cycles elapsed from the reference time (see left-top panel in Fig.~\ref{fig:tstars}). $\Delta T_{\text{NOD}}$ represents the difference between the predicted passage from the ascending node $T_{\text{NOD},predict}=T_{\text{NOD},1998}+N P_{orb_{1998}}$ and the reference value $T_{\text{NOD},1998}$, where the integer $N$ is the number of orbital cycles elapsed between two different $T_{\text{NOD}}$. We started by fitting the data with a quadratic function: 
\begin{eqnarray}
\label{eq:fit_tstar}
\Delta T_{\text{NOD}} = \delta T_{\text{NOD},1998} + N\, \delta P_{orb_{1998}}+0.5\,N^2\, \dot{P}_{orb}P_{orb_{1998}},
\end{eqnarray} 
where the correction to the adopted time of passage from the ascending node, $\delta T_{\text{NOD},1998}$, the correction to the orbital period, $\delta P_{orb_{1998}}$, and the orbital-period derivative, $\dot{P}_{orb}$, are the fit parameters. Statistically speaking, the fit is not acceptable, since a $\chi^2=429.5$ (for 4 degrees of freedom) corresponds to a probability of obtaining a larger $\chi^2$ of $\sim2.4\times 10^{-6}$ (largely below the conventional accepted 5\% threshold). In Tab.~\ref{tab:par_fit_orb} we reported the updated orbital parameters inferred from the best-fitting model. To take into account the large value of the reduced $\tilde{\chi}^2$ obtained from the fit, we rescaled the uncertainties of the fit parameters by the quantity $\sqrt{\tilde{\chi}^2}$.
\begin{table}
\begin{tabular}{l | c  }
Parameters             & value \\
\hline
Ascending node passage $T_{\text{NOD}}$ (MJD) & 50914.79452(1)\\
Orbital period $P_{orb}$ (s) & 7249.15651(9)\\
Orbital period derivative $\dot{P}_{orb}$ (s s$^{-1}$) & $3.6(4)\times 10^{-12}$\\
$\chi^2$/d.o.f. &429.5/4\\
\hline
\end{tabular}
\caption{\saxj{} best-fitting orbital parameters derived combining the seven outbursts observed between 1998 and 2015. All parameters are referred to the 1998 outburst, with $T_{\text{NOD}, 1998}=50914.794528(2)$ (MJD) and $P_{orb, 1998}=7249.156444(23)$ s \citep{Burderi09}. Uncertainties are reported at $1\sigma$ confidence level. Uncertainties are scaled by a factor $\sqrt{\tilde{\chi}^2}$.}
\label{tab:par_fit_orb}
\end{table}
We investigated the presence of an orbital period second derivative \citep[as suggested by][]{Patruno12a} by adding a cubic term in Eq.~\ref{eq:fit_tstar}. The fit with the updated model is statistically unacceptable, with $\chi^2_{\nu}=173.3$ for 3 d.o.f. (p-value probability of $\sim1\times 10^{-6}$). Moreover, with a null-hypothesis probability of $\sim0.07$, we note that the $\chi^2$ variation caused by introducing the new component ($\Delta \chi^2\sim 256$ for 1 d.o.f) is not statistically significant for the specific model. We therefore ignored it for the following discussion.
As we can see from the residuals reported in the left-bottom panel of Fig.~\ref{fig:tstars}, the farthest outlier (more than $30\sigma$) from the best-fitting quadratic model corresponds to the 2011 outburst ($\sim59000$ orbital cycles) from which \citet{Patruno12a} inferred the time second derivative of the orbital period. Removing this outburst lowers the $\chi^2_\nu$ to $\sim9.6$ (with 3 d.o.f.), consistent with a constant orbital period derivative of $\dot{P}_{orb}=3.56(6)\times 10^{-12}$ s s$^{-1}$. Both values of $\dot{P}_{orb}$ inferred from the analysis (with or without the 2011 outburst) are compatible within errors with the estimates reported by \citet{diSalvo08} and \citet{Hartman08} for the orbital evolution of the source up to the 2005 outburst, and with the estimates of \citet{Hartman09b} and \citet{Burderi09} up to the 2009 outburst.

The origin of the observed $\dot{P}_{orb}$ is still not fully understood, yet different possible mechanisms have been proposed over the years \citep[see e.g.,][]{diSalvo08, Hartman08,Burderi09, Patruno12a}. However, there is consensus on the fact that conservative mass transfer is not compatible with the observed value of $\dot{P}_{orb}$ for \saxj{}. This can be easily demonstrated by estimating the mass-loss rate from the secondary as a function of the observed orbital period derivative. Combining Kepler's third law with the condition for mass transfer via Roche-lobe overflow ($\dot{R}_{L2}/R_{L2}=\dot{R}_2/R_{2}$, where the $R_{L2}$ and $R_{2}$ are the Roche-lobe radius and radius of the secondary, respectively), we can write the averaged secondary mass-loss rate as \citep[see][for further details on the derivation of the expression]{Burderi2010a}:
\begin{eqnarray}
\label{eq:m2_pdot}
\dot{m}_{2}=1.2 \times (3n-1)^{-1}\,m_{2,0.14\odot{}}\Big(\frac{\dot{P}_{orb,-12}}{P_{orb,2h}}\Big)\times 10^{-9} \text{M}_{\odot}\, \text{yr}^{-1} ,
\end{eqnarray}
where $n$ is the index of the mass-radius relation of the secondary R$_2\propto$ M$_2^n$, $m_{2,0.14\odot{}}$ is the mass of the companion star in units of 0.14 M$_{\odot}$ \citep[from the mass function it can be inferred that $m_2 \leq 0.14$ M$_\odot$ at 95\% confidence level, see e.g.,][]{Chakrabarty1998a}, $\dot{P}_{orb,-12}$ is the orbital-period derivative in units of $10^{-12}$ s s$^{-1}$, and $P_{orb,2h}$ is the orbital period in units of 2 hours (appropriate for \saxj{} since $P_{orb}\approx 2.01$ hours). Since the transferred mass is lost by the companion star, the quantity on the right side of Eq.~\ref{eq:m2_pdot} must be negative, which implies a secondary mass-radius index $n<1/3$. Assuming a fully convective companion star ($n=-1/3$), we find $\dot{m}_{2}\simeq-2\times 10^{-9} \text{M}_{\odot} \text{yr}^{-1}$. The conservative mass transfer scenario implies that mass transferred during the outburst must be completely accreted by the NS, while no mass is accreted or lost during quiescence phases. To verify whether the value inferred from Eq.~\ref{eq:m2_pdot} is compatible with a conservative scenario, we extrapolated the companion averaged mass-loss rate from the observed flux of the source. Combining the observations of the source during the outburst phases shown in the past seventeen years of monitoring, and taking into account that the source spends on average 30 days in outburst every 2.5 years, we can infer the averaged mass-loss rate $\dot{m}_{2,obs}\sim2\times 10^{-11} \text{M}_{\odot} \text{yr}^{-1}$. The discrepancy of two order of magnitude between the mass-loss rate values strongly suggests that the observed orbital period derivative is not compatible with a conservative mass-transfer scenario.

\subsubsection{Gravitation quadrupole coupling} As noted by \citet{Hartman08}, the large orbital period derivative observed in \saxj{}, as well as its orbital parameters, are very similar to those of a small group of black widow millisecond pulsars, such as PSR B1957$+$20 \citep{Arzoumanian1994a, Applegate1994a}, and PSR J2051$-$0827 \citep{Doroshenko2001a, Lazaridis2011a, Shaifullah2016a}. Similar orbital period variations have been observed in the red-back system PSR J2339$-$0533 \citep{Pletsch2015a}, in the transitional red-back system PSR J1023$+$0038 \citep{Archibald2013a}, and in the LMXB system EXO 0748$-$676 \citep{Wolff09}. 

Orbital period variations observed in these systems have been interpreted in terms of gravitational quadrupole coupling (GQC), i.e. gravitational coupling between the orbit and the variations of the quadrupole moment of the magnetically active companion star \citep{Applegate1992a, Applegate1994a}. Applegate's GQC model proposes that magnetic fields can vary the mass distribution of an active star by transitioning different states of fluid hydrostatic equilibrium. The variable deformation of the companion is then explained as the result of the star angular momentum redistribution generated likely by a magnetic torque that causes cyclic spin-up and spin-down of the companion's outer layers. Variations of the companion oblateness will then be communicated on short (almost dynamical) timescales by gravity to the orbit inducing orbital period changes. An increase in the quadrupole moment will cause the orbit to shrink (negative orbital period derivative), vice versa a decrease in the quadrupole moment will result in a widening of the orbit (positive orbital period derivative). We can express the variation of the orbital period in terms of the variable quadrupole moment $\Delta Q$ as:
\begin{eqnarray}
\frac{\Delta P_{orb}}{P_{orb}}= -9\frac{\Delta Q}{m_2 a^2},
\label{eq:ap1}
\end{eqnarray}
where $m_2$ is the companion mass and $a$ is the orbital separation \citep{Applegate1994a}. The variation of the quadrupole moment of the companion can be related to the change in the angular velocity ($\Delta \Omega$) of the outer layers caused by the transfer of angular momentum ($\Delta J$) by the relation:
\begin{eqnarray}
\Delta Q= \frac{2\,m_s\, R^5_2}{9\,G\, m_2} \Omega \, \Delta \Omega,
\label{eq:ap2}
\end{eqnarray}
where $\Omega$ is the star angular velocity, and $R_2$ and $m_s$ represent the radius and the mass of a thin outermost shell of the companion star, respectively. Combining Eq.~\ref{eq:ap1} and Eq.~\ref{eq:ap2}, and assuming that $\Omega$ is almost synchronous with respect to the orbital angular velocity, we obtain the variable angular velocity required to produce orbital period changes $\Delta P_{orb}$ \citep{Applegate1994a}
\begin{eqnarray}
\frac{\Delta \Omega}{\Omega}= \frac{G\,m_2^2}{2R_2^3\,m_s} \left(\frac{a}{R_2}\right)^2\left(\frac{P_{orb}}{2\pi}\right)^2 \frac{\Delta P_{orb}}{P_{orb}}.
\label{eq:ap3}
\end{eqnarray}
To investigate this scenario we attempted to describe the data reported in Fig.~\ref{fig:tstars} assuming the following prescription for the orbital period:

\begin{eqnarray}
\label{eq:ASmod}
\Delta T_{\text{NOD}} &=& \delta T_{\text{NOD,1998}} + N\, \delta P_{orb_{1998}}\\\nonumber
&&+A\sin\left(\frac{2\pi}{P_{mod}} N P_{orb_{1998}} +\psi \right),
\end{eqnarray} 
where $A$ represents the amplitude of the oscillation shown by the $\Delta T_{\text{NOD}}$, $P_{mod}$ is the modulation period and $\psi$ is modulation phase. We obtained the best fit ($\chi^2_{\nu}=26.7$ for 3 d.o.f.) for $ \delta P_{orb_{1998}}=9.3(3)\times 10^{-4}$ s, $P_{mod}=21.1(1.8)$ yr and $A=10.1(7)$ s. It is worth noting that the best-fit residuals (Fig.~\ref{fig:tstars}, bottom panel) improved with respect to the previous model, however, the reported $\chi^2_\nu$ value indicates a statistically unacceptable fit. Starting from the amplitude of the oscillation $A$ measured from the fit of $\Delta T_{\text{NOD}}$, we can infer the amplitude of the orbital period modulation via the relation
\begin{eqnarray}
\label{eq:ap4}
\frac{\Delta P_{orb}}{P_{orb}}=2\pi \frac{A}{P_{mod}}=0.95(7)\times 10^{-7}.
\end{eqnarray}
According to \citet{Applegate1994a}, the variable part of the luminosity $\Delta L$ required to produce orbital period change $\Delta P_{orb}$ is given by 
\begin{eqnarray}
\Delta L\simeq \frac{\pi}{3} \frac{G\,m_2^2}{R_2 P_{mod}} \left(\frac{a}{R_2}\right)^2 \frac{\Delta \Omega}{\Omega} \frac{\Delta P_{orb}}{P_{orb}}.
\label{eq:ap5}
\end{eqnarray}
To quantify $\Delta L$ we made the following assumptions: i) the thin shell that differentially rotates with respect to the companion star has a mass $m_s\simeq 0.1 m_2$ \citep{Applegate1994a}; ii) we approximate the companion star Roche-Lobe radius as $R_{L2}\simeq 0.462\, [q/(1+q)]^{1/3} a$, valid for mass ratio $q\leq0.8$ \citep{Paczynski71}; iii) the conversion efficiency of internal luminosity into mechanical energy to power the shell oscillations in the gravitational potential of the companion is of the order of 10\% \citep{Applegate1992a}; iv) combining Roche-Lobe overflow mass transfer condition ($R_2 \simeq R_{L2}$) with the binary mass function, we can express the companion mass radius as $R_2\simeq 0.37\, m_{2,\odot{}}^{1/3}\, P_{orb,2h}$ R$_{\odot}$, where $m_{2,\odot{}}$ represents the companion mass in units of solar masses and $P_{orb,2h}$ is the orbital binary period in units of 2 hours. Using the aforementioned assumption in Eq.~\ref{eq:ap3} and Eq.~\ref{eq:ap5}, we find the internal energy required to power the GQC mechanism:
\begin{equation}
L_{\text{GQC}}=1.5\times 10^{32} m_{1,\odot} q^{1/3}(1+q)^{4/3}P_{orb,2h}^{-2/3}\frac{A^2}{P_{mod,yr}^3}\, \text{erg/s},
\label{eq:ap5b}
\end{equation} 
where m$_{1,\odot{}}$ represents the NS mass in units of solar masses and $P_{mod,yr}$ is the modulation period in units of years.
Substituting the parameters of \saxj{} in Eq.~\ref{eq:ap5b} (in particular a companion mass of 0.047 M$_\odot$ obtained assuming a NS mass of 1.4 M$_{\odot}$, and a binary inclination of $65^\circ$ derived from the modelling of the reflection component; Di Salvo et al., in prep.), the companion star must have a source of energy capable of supplying an internal luminosity $L_{\text{GQC}}\sim 10^{30}$ erg s$^{-1}$. 

\noindent
Another important aspect that should be discussed is the magnetic field strength required to produce the orbital period variation observed in the system. Starting from the formula reported by \citet{Applegate1992a}, the mean subsurface field can be expressed as:
\begin{eqnarray}
B\simeq 3.4\times 10^{4} (m_{1,\odot{}}+m_{2,\odot})^{1/3}P_{orb,2h}^{-3/2} A^{1/2} P_{mod,yr}^{-1} \,\text{Gauss,}
\end{eqnarray}
Substituting the results obtained for \saxj{}, we estimate a mean subsurface field $B\simeq6\times 10^3$ G. Interestingly, similar magnetic field values have been reported by \citet{Reiners2012a} from the observation of isolated brown dwarfs and low-mass stars. In the following we analyse all the possible source of energy that could power the GQC oscillations.
\subsubsection*{Source of energy to power the GQC mechanism}

\begin{itemize}
\item \textit{Nuclear energy.} To determine the internal luminosity available in the core of the very low massive and likely old companion star of \saxj{}, we simulated the evolution of such a star by means of stellar evolutionary code \citep[\textsc{ATON}, for more details see e.g.,][]{Ventura1998a, Ventura2008a, Tailo2015a, Tailo2016a}. Starting from a 1.0 M$_{\odot}$ star with solar chemistry, we made it loose mass at constant rate ($\dot M$ = $1.0\times10^{-9}$ M$_{\odot}$/yr), as it evolved during the hydrogen burning phase, down to the point it reached the mass of 0.047 M$_{\odot}$. Subsequently we left it evolve until we were able to do so. For a companion mass of $m_2=0.047$ M$_{\odot}$, for which nuclear burning is not active anymore, we estimated a luminosity $L_2\simeq10^{-4.5} L_{\odot}\simeq1.2\times 10^{29}$ erg s$^{-1}$ \citep[similar results can be obtained following the stellar properties reported in literature, see e.g.,][]{Chabrier1997a}, almost an order of magnitude fainter than the source of energy required to power the GQC effects previously described.
\item \textit{Companion star irradiation}. The optical counterpart of \saxj{} (V4584 Sagittarii)  shows spectral properties characteristic of an evolved (mid-to-late type) star \citep{Roche1998a}, likely reflecting a very low-mass (irradiation-bloated) brown dwarf \citep[see e.g.,][]{Bildsten2001a}. High-time resolution CCD photometry observations of the optical counterpart during the X-ray quiescence phase of \saxj{} allowed to infer an optical flux of $V\sim 21.5$ mag \citep{Homer2001a}. The lack of ellipsoidal variations in the flux emission combined with the observed sinusoidal binary modulation have been explained by \citet{Homer2001a} as the result of the irradiation of the companion star face by a quiescence irradiating X-ray flux ($L_{irr}\sim 10^{33}$ erg s$^{-1}$) compatible with low-level mass transfer driven by gravitational radiation losses, but incompatible with the X-ray luminosity of the source observed in quiescence, $\sim10^{31.5}$ erg s$^{-1}$ \citep{Campana2002a}. Alternatively, \citet{Burderi2003a} \citep[see also][]{Campana2004a} proposed that the observed optical flux might originate from the illumination of the companion by the rotation-powered pulsar emission during quiescence, which switches on once the magnetospheric radius exceeds the light cylinder radius. The power released by the active magneto-dipole emitter can be expressed as $L_{\textsc{PSR}}=2/(3c^3)\mu^2\omega^4\simeq1.6\times 10^{34} I_{45}\,\nu_{401}\,\dot{\nu}_{-15,\textsc{SD}}$ erg s$^{-1}$, where $I_{45}$ represents the NS moment of inertia in units of $10^{45}$ g cm$^2$, $\mu_{401}$ is the NS spin frequency in units of 401 Hz and $\dot{\nu}_{-15,\textsc{SD}}$ represents the spin-down frequency derivative in units of $10^{-15}$ Hz/s. Substituting the value $\dot{\nu}_{-15,\textsc{SD}}\simeq2.5$ (see Sec.~\ref{sec:spin} for more details), we estimate a luminosity $L_{\textsc{PSR}}\simeq4\times 10^{34}$ erg s$^{-1}$. In the hypothesis of an isotropic irradiation, only a fraction $f=(1-\cos \theta)/2$ is intercepted by the companion star, where $\theta$ represents half of the angle subtended by the secondary as seen from the NS and it relates to the binary system by the expression $\tan \theta = R_2/a$, with $R_2$ and $a$ being the companion star radius and the orbital separation, respectively. Since the companion star fills its Roche-lobe, we can approximate $R_2$ with $R_{L2}\simeq0.462\,[q(1+q)]^{1/3}a$ \citep{Paczynski71}. Assuming $m_1=1.4$ M$_{\odot}$ and $m_2=0.047$ M$_{\odot}$, we obtain that only a fraction $f\simeq0.006$ of the pulsar luminosity will be intercepted and reprocessed by the companion star, corresponding to a luminosity of $\sim2.4\times 10^{32}$ erg s$^{-1}$. Adopting an efficiency factor of the order of 10\% \citep[see][for more details]{Applegate1994a} to convert the irradiation energy into mechanical energy, we estimate that the radio pulsar could potentially provide enough energy to power the observed GQC effects. However, as described by \citet{Applegate1992a}, the cyclic magnetic activity that generates the variable quadrupole moment needs to be powered from the inner regions of the star. Therefore, it remains to be investigated whether the irradiation energy impinging on the companion star can be converted to internal energy. The total energy generated by the rotating magnetic dipole and intercepted by the companion, as described above, is 
$\sim 2\times 10^{32}$ erg s$^{-1}$, characterised by a quite uncertain spectral distribution. Broadly speaking, a consistent fraction of this radiation is emitted in the form of Ultra-Low Frequency Electromagnetic Radiation (ULF, hereafter) at the spin frequency of the NS, namely $\sim400$ Hz. These frequency is well below the plasma frequency of the companion photosphere, $\nu_p \simeq 7\times 10^{6} \rho^{1/2}(X+Y/2)^{1/2}$ GHz, where $\rho$ is the density of the photosphere, and $X$ and $Y$ represent the hydrogen and helium mass fraction, respectively. In this situation plasma behaves has a metal almost entirely reflecting the impinging radiation \citep[see e.g.,][]{Stenson2017a}, which is therefore unable to effectively penetrate the external layers of the companion and power the GQC mechanism. In line with the Goldreich and Julian model \citep[see][\citealt{Hirotani2006a} and references therein]{Goldreich1969a}, the remaining power, still of the order of $\times 10^{32}$ erg s$^{-1}$, is emitted in the form of high energy gamma rays and electron-positron relativistic pairs. According to \citet{Hameury1996a}, this energetic radiation is capable to penetrate up to a column depth of $\sim 100$ g cm$^{-2}$, with corresponds to a physical depth well below few kilometres, i.e. less than $1\times 10^{-4}$ of the Roche-lobe filling companion. This penetration is by far to shallow to power the GQC mechanism in which the displacement of the shell of matter responsible for quadrupole-moment variation is of the order of a tenth of the stellar radius \citep{Applegate1992a}. Furthermore, we consider the irradiation of the companion star face as the result of the quiescence irradiating X-ray flux. According to \citet{Hameury1996a, Hameury1997a} and \citet{Podsiadlowski1991a}, the X-ray penetration on the companion star is even less than $100$ g cm$^{-2}$. All this seems to suggest that the internal energy required to power the GQC mechanism cannot be supplied by external irradiation of the companion star.

\item \textit{Tidal dissipation}. Another possible mechanism to transfer energy to the companion star is via tidal dissipation. The key ingredient for this process is the asynchronism between the binary system and the companion star. The same magnetic activity invoked to explain the variable quadrupole moment of the companion star could contribute to a torque that slows down the spin of the secondary by means of a magnetic-braking-like mechanism supported by mass-loss from irradiation driven winds powered by the pulsar emission intercepted by the companion. The torque holds the companion star out of synchronous rotation with respect to the binary system, generating a tidal torque and consequently tidal dissipation. Following Eq.~21 in \citet{Applegate1994a}, to obtain a tidal luminosity of the order $10^{30}$ erg s$^{-1}$ (luminosity required to power the orbital period changes previously described), the following mass-loss rate is required:
\begin{eqnarray}
\dot{m}_{T}=1.16\times 10^{-10}\left(\frac{a}{l}\right)^2L_{\text{30}}^{1/2}\,t_{syn,4}^{-1/2}\,I_{2,51}^{1/2}\,\,\, \text{M$_{\odot}$ yr$^{-1}$,}
\label{eq:ap6}
\end{eqnarray} 
where $(a/l)$ represents the ratio between the binary separation and the lever arm of the mass ejected from the companion star, $L_{30}$ is the tidal luminosity in units of $10^{30}$ erg s$^{-1}$ and $I_{2,51}=0.1 m_2 R_{L2}^2$ is the donor moment of inertia defined in units of $10^{51}$ g cm$^2$. The parameter $t_{syn,4}$ represents the tidal synchronisation time in units of $10^4$ years that is defined as:
\begin{eqnarray}
t_{sys, 4}=\frac{0.65}{\mu_{12}}\left(\frac{R_{L2}}{R_2}\right)^6(1+q)^2\frac{m_{2,\odot}}{R_{2,\odot}},
\label{eq:ap7}
\end{eqnarray}
where $\mu_{12}=3\times 10^{12}\,L_{T,\odot}^{1/3}\,R_{2,\odot}^{-5/3}\,m_2^{2/3}$ g cm$^{-1}$ s$^{-1}$ is the mean dynamic viscosity with L$_{T,\odot}$ and R$_{2,\odot}$ representing the companion star luminosity and radius, respectively, all rescaled in solar units. Considering $R_2\simeq R_{L2}$ and $m_{2,\odot}=0.047$ we find that for $L_T\sim10^{30}$ erg s$^{-1}$ the synchronisation time is approximately $3.5\times10^3$ years. Substituting these values into Eq.~\ref{eq:ap6} and assuming a magnetic lever arm $l\simeq0.5 a$ \citep[in analogy with][]{Applegate1994a}, we estimate the tidal mass-loss to be of the order $\sim 1.5\times10^{-9}$ M$_{\odot}$ yr$^{-1}$. Interestingly, the estimated mass-loss rate is similar to the value inferred from the analysis of the NS spin-up during its 2002 outburst \citep[][see, however \citealt{Hartman08} who suggest that no NS spin period derivative is observable during that outburst]{Burderi06}, and from the global parabolic trend interpreted as the orbital period derivative. In fact, a companion mass-loss rate as the one estimated must have an influence on the orbital period evolution of the system, that it is not taken into account by the GQC mechanism described by \citet{Applegate1994a}, though the consequent orbital expansion is clearly mentioned in their work. Following \citet{Burderi09}, the orbital period derivative caused by GR angular momentum and matter losses can be expressed as follows:
 \begin{eqnarray}
\dot{P}_{orb,-12}&=&-1.38\,m_{1,\odot{}}^{5/3}q(1+q)^{-1/3}P_{orb,2h}^{-5/3}+\\ \nonumber
&&+0.648\,m_{1,\odot{}}^{-1}q^{-1}P_{orb,2h}\,g(\beta, q, \alpha)\,\dot{m}_{-9},
\label{eq:orb}
\end{eqnarray}   
where $q$ is the mass ratio between the companion and the compact object, $g(\beta, q, \alpha)=1-\beta q-(1-\beta)(\alpha+q/3)/(1+q)$ reflects the angular momentum losses because of mass lost from the system with $\beta$ being the fraction of the mass lost by the companion star that is accreted onto the NS and $\alpha=l_{ej}P_{orb}\,(m_1+m_2)^2/(2\pi a^2 m_1^2)$ the specific angular momentum of the matter leaving the system ($l_{ej}$) in units of the specific angular momentum of the companion star located at a distance $r_2$ from the center of mass of the system with an orbital separation $a$. To take into account the correlation between the GQC effects, the ejected matter required for the tidal dissipation process and its effect on the orbital evolution, we modified the model reported in Eq.~\ref{eq:ASmod} as follows:
\begin{eqnarray}
\label{eq:ASmod2}
\Delta T_{\text{NOD}} &=& \delta T_{\text{NOD,1998}} + N\, \delta P_{orb_{1998}}\\\nonumber
&&+\frac{1}{2}P_{orb_{1998}}\dot{P}_{orb}+A\sin\left(\frac{2\pi}{P_{mod}} N P_{orb_{1998}} +\psi \right),
\end{eqnarray} 
where $\dot{P}_{orb}$ is fixed to the value obtained from Eq.~\ref{eq:orb} when inserting the mass-loss rate estimated from Eq.~\ref{eq:ap6}. This new model allows us to fit the GQC effects properly taking into account the response of the binary system caused by the ejection of the matter required to power the GQC mechanism. In the right-top panel of Fig.~\ref{fig:tstars} we show a possible application of this model to \saxj{}. More specifically, the best-fit model reported here has been obtained assuming i) a magnetic lever arm $l\simeq 0.5a$ and ii) that matter is ejected from the system with a specific angular momentum proportional to its distance from the binary centre of mass along the line connecting the two component of the system. This assumption implies that, during the magnetic-braking-like mechanisms, the matter remains attached to the field lines for a time interval much shorter than the companion orbital period. The obtained best-fit corresponds to the following parameters: $ \delta P_{orb_{1998}}=4(0.7)\times 10^{-4}$ s, $P_{mod}=14.9(2.7)$ yr, $A=3.6(6)$ s and $\dot{P}_{orb}=2.3\times 10^{-12}$ s s$^{-1}$, corresponding to a mass-loss rate of $\sim8\times 10^{-10}$  M$_{\odot}$ yr$^{-1}$. Although statistically unacceptable ($\chi^2_\nu$=41.6 for 3 d.o.f.), the distribution of the best-fit residuals (Fig.~\ref{fig:tstars}, right-bottom panel) suggests that the model is potentially capable to describe the orbital evolution of \saxj{}. However, it is worth noting that at the moment the data sample is too small to allow a self-consistent determination of all the model parameters, such as the magnetic field lever arms or the specific angular momentum of the matter ejected from the system. More outbursts of the source are therefore required to confirm the presence of periodic oscillations of the orbital period; in alternative these may be random oscillations of the orbital period around a global parabolic trend, possibly caused by random fluctuations of a non-conservative mass transfer rate.

\end{itemize}

\subsubsection{Non-conservative mass transfer scenario} \citet{diSalvo08} and \citet{Burderi09} proposed an alternative scenario to explain the large orbital period derivative, that invokes a non-conservative mass transfer from the companion star to the NS. The idea is that even during the quiescence phases the companion star overflows its Roche-lobe, but the transferred matter is ejected from the system rather than falling onto the NS surface. Using Eq.~5 from \citet{diSalvo08}, the secular orbital period derivative of the system driven by emission of gravitational waves and mass loss can be expressed as:
\begin{eqnarray}
\dot{P}_{orb}&=& [-0.138m_{1,\odot{}}^{5/3}\, q_{0.1}(1+0.1q_{0.1})^{-1/3}P^{-5/3}_{orb,2h}\\\nonumber
&+&6.84m_{1,\odot{}}^{-1}\,q_{0.1}^{-1}P_{orb,2h}\,g(\beta, q, \alpha)\,\dot{m}_{2,-9}]\times 10^{-12} \,s s^{-1},
\end{eqnarray}
where $q_{0.1}$ is the mass ratio in units of 0.1.
\begin{figure}
\centering
\includegraphics[width=0.48\textwidth]{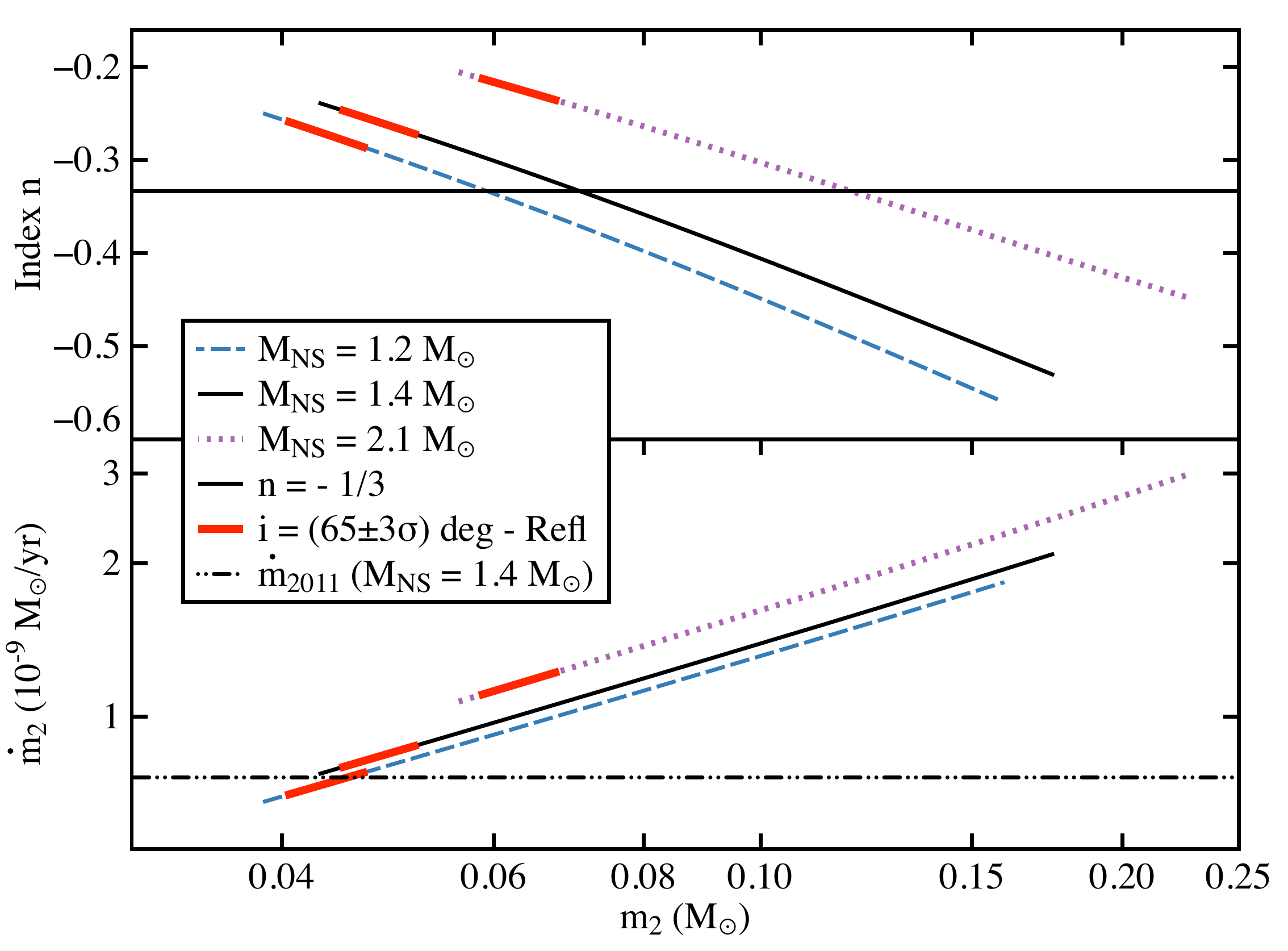}
\caption{Top panel - Mass-radius index versus mass of the companion star evaluated assuming a highly non-conservative ($\beta\sim 0.01$) secular mass-transfer scenario for the binary system, with mass ejected at the inner Lagrangian point $L_1$ (see the text for details on the model). Blue dashed line, black solid line and purple dotted line have been estimated by assuming a NS mass of $1.2$ M$_{\odot}$, $1.4$ M$_{\odot}$  and  $2.1$ M$_{\odot}$, respectively. The horizontal black line represents the mass-radius index $n=-1/3$ characterising fully convective stars.  Bottom panel - Mass-loss rate versus mass of the companion star predicted by adopting the finding reported in the top panel. The horizontal dot-dot-dashed line represents the inferred bolometric mass-accretion rate extrapolated from the peak of the 2011 outburst of the source.
Mass values of the companion star mass have been estimated from the binary mass function assuming inclination angles between $90^\circ$ and $15^\circ$\citep[corresponding to the 95 per cent confidence level, see e.g.,][]{Chakrabarty1998a}. Red-solid segments highlight the values of $n$ and $\dot{m}_2$ corresponding to the orbital inclination ($i=65^\circ\pm9^\circ$) derived from the modelling of the reflection spectrum observed during the latest outburst of the source.}
\label{fig:index_mass}
\end{figure}
Substituting Eq.~\ref{eq:m2_pdot} in the previous expression and using the orbital period derivative determined from the timing analysis ($\dot{P}_{orb}\sim3.6\times 10^{-12}$ s/s), we derive a relation between the two masses of the objects forming the binary system ($m_1$ and $m_2$) and the companion mass-radius index ($n$). Top panel of Fig.~\ref{fig:index_mass} shows the mass-radius index as a function of the companion mass assuming three different NS mass values ($1.2$ M$_{\odot}$ cyan, $1.4$ M$_{\odot}$ black and  $2.1$ M$_{\odot}$ purple). The three functions are calculated under the assumptions that i) the mass ejected from the system leaves the system with the specific angular momentum at the inner Lagrangian point $\alpha_{L1}=[1-0.462(1+q)^{2/3}q^{1/3}]^2$ and ii) the fraction of mass lost from the companion and accreted onto the NS is $\beta\sim0.01$, roughly corresponding to the ratio between the times of X-ray activity and quiescence phase occurred since the source discovery. The range of masses displayed for the companion star have been determined from the mass function estimated from the updated ephemeris of the X-ray pulsar and assuming inclination angles between $90^\circ$ and $15^\circ$ \citep[corresponding to the 95\% confidence level, see e.g.,][]{Chakrabarty1998a}. Moreover, the horizontal line represents the mass-radius index $n=-1/3$, typically used to describe degenerate or fully convective stars. Taking into account the binary inclination ($i=65^\circ\pm9^\circ$ with a $3\sigma$ confidence level uncertainty) derived from the spectral properties of the reflection spectrum modelled from the \xmm{} and \nustar{} observation of the latest outburst (Di Salvo et al. in prep.), we predict the mass and the mass-radius index of the companion mass (red solid lines in Fig.~\ref{fig:index_mass}) to be $m_2=(0.04-0.047)$ M$_\odot$ and $n=-(0.29-0.27)$, $m_2=(0.044-0.052)$ M$_\odot$ and $n=-(0.27-0.26)$, $m_2=(0.062-0.068)$ and M$_\odot$ $n=-(0.24-0.22)$ for a NS mass of $1.2$ M$_{\odot}$, $1.4$ M$_{\odot}$  and  $2.1$ M$_{\odot}$, respectively. It is worth noting that, even though the companion mass values are well below 0.1 M$_{\odot}$, the mass-radius index appears slightly larger than $n=-1/3$, predicted for star that became fully convective \citep[see e.g.,][]{King88, Verbunt93}. This discrepancy could reflect the different secular evolution that the companion star may experience within the binary system compared to a standard isolated star. Nonetheless, the effect of the direct X-ray and magnetic dipole irradiation of the companion star during the outburst activity and the quiescence phase, respectively, could likely modify the mass-radius hydrostatic equilibrium relation of the star. Bottom panel of Fig.~\ref{fig:index_mass} shows the predicted companion star mass-transfer rate estimated by substituting $m_1$, $m_2$ and $n$ reported in the top panel into Eq.~\ref{eq:m2_pdot}. As a reference, we show the mass accretion rate measured during the peak of the 2011 X-ray outburst of the source, inferred assuming a source distance of $3.5$ kpc.  
In line with \citet{diSalvo08} and \citet{Burderi09}, we suggest that the up-to-date measured value of the orbital period derivative (reflecting a long-term evolution of seventeen years) is compatible with a highly non-conservative mass transfer scenario where almost 99\% of the matter transferred from the companion star is ejected from the system and not directly observed. 

\subsubsection*{Radio-ejection mechanism}
The proposed mechanism to expel the accreting matter from the system involves radiation pressure of the magneto-dipole rotator that should power the NS during quiescence phases \citep[see e.g.,][]{Burderi2003a, diSalvo08}. Over the years several indirect hints that in \saxj{} the radio pulsar mechanism turns on during quiescence phases have been collected, e.g., the presence of an over-luminous optical counterpart of the source \citep{Homer2001a} interpreted by \citet{Burderi2003a} as the spin-down luminosity of the magneto-dipole rotator reprocessed by the companion star; the secular spin-down evolution of the NS spin showing derivative values typical of millisecond radio pulsars \citep{Hartman09b, Patruno12a}; detection of a possible gamma-ray counterpart of the source as observed by \textit{Fermi-LAT} during the quiescence phases of \saxj{} \citep{de-Ona-Wilhelmi2016a, Xing2015a}. Nonetheless, although carefully searched \citep{Burgay2003a, Patruno2016a}, no direct observation of the radio pulsar activity of the source has been observed to date \citep[at this moment the only known AMXP observed as rotation-powered pulsar is IGR J18245-2453,][]{Papitto2013b}. In analogy with the black widows in the stage of the so-called \emph{star-vaporising pulsars} or \emph{hidden} millisecond pulsars \citep[e.g.,][]{Tavani1991a}, \citet{diSalvo08} suggested that the radio emission from the rotation powered pulsar is completely blocked by material engulfing the system likely generated from the interaction between the pulsar radiation pressure and the mass overflowing from the companion star (the so-called hidden black widow scenario). X-ray outbursts would then be triggered by temporary increase of pressure of the overflowing matter sufficient to overcome the radiation pressure of the rotation-powered pulsar. Discrepancies observed around the almost constant secular expansion of the orbital period (Fig.~\ref{fig:tstars}, left-bottom panel) could then reflect stochastic fluctuations of the radiation pressure regulating the mass-loss rate, or of the mass transfer rate itself.

\citet{Hartman09b} verified the energetic feasibility of the aforementioned scenario showing that the spin-down luminosity of \saxj{} in quiescence (estimated from the observed long-term spin down of the pulsar) could drive a mass loss from the companion of the order of $10^{-9}$ M$_{\odot}$ yr$^{-1}$, compatible with our findings reported above. However, the authors raised doubts on the possibility to preserve the accretion disk during the activity of the particle wind, hence the possibility to switch from rotation-powered to accretion-powered pulsar.

\section{conclusions}
In this paper, we presented an analysis of the \xmm{} and \nustar{} observations collected during the 2015 outburst of \saxj{}. From the timing analysis of the two datasets, we determined an updated set of ephemeris of the source. We investigated the time evolution of the spin frequency by means of pulse phase timing techniques applied to the fundamental and the second harmonic components used to describe the coherent signal. We reported significant spin frequency derivative values from the analysis of the fundamental component of both datasets. Moreover, for \xmm{}, we observed a clear mismatch between the time evolution of the fundamental and its second harmonic component, likely connected to the phase-jumping phenomenon responsible of the strong timing noise present in this source. We extended the study of the secular spin evolution combining the frequency estimates from the previous six outbursts observed between 1998 and 2011 with the value estimated from the analysis of the second harmonic component during the latest outburst in 2015. From the best-fitting linear model, we estimated a long-term spin frequency derivative of $\sim-1.5\times 10^{-15}$ Hz s$^{-1}$, compatible with a NS magnetic field of $B\sim2\times10^{8}$ G. Furthermore, we estimated possible corrections to the magnetic field taking into account the effect of mass accretion onto the NS during the outbursts.
Combining the seven monitored outbursts of the source, we confirmed the fast orbital expansion of \saxj{} characterised by an average orbital period derivative of $\sim3.6\times 10^{-12}$ s s$^{-1}$, not compatible with a conservative mass-transfer scenario driven by GR. We investigated the possibility to explain the observed orbital evolution with the gravitational quadrupole coupling model \citep{Applegate1992a}. We found that, for the specific case of \saxj{}, such a scenario would require a strong magnetic field (several kG) powered by a tidal dissipation mechanism acting on the companion star. Moreover, this scenario would also imply a large fraction of ejected matter that would strongly influence the orbital evolution of the system. We suggest that under specific conditions this scenario is compatible with the observed orbital evolution of \saxj{}. 
Finally, we also discuss the large and almost stable long-term orbital period derivative characterising \saxj{} in terms of a highly non-conservative mass transfer scenario, where a large fraction ($\sim99\%$) of the mass transferred from the companion star is ejected from the inner Lagrangian point as a consequence of the irradiation from the magneto-dipole rotator during the quiescent phase of the system (\textit{radio-ejection} model). In both cases, therefore, the mass transfer in the system has to be highly not conservative. The main difference is that in the first scenario matter is expelled along the magnetic field lines of the companion with a lever arm of half the orbital separation along the line connecting the two stars with the specific angular momentum with respect to the center of mass of the system; this is needed in order to power GQC oscillations via tidal dissipation. On the other hand, in the second scenario matter is expelled directly at the inner Lagrangian point, and the variations of the orbital period with respect to the global parabolic trend are caused by random fluctuations of the mass transfer rate. Indeed it is not excluded that the \textit{radio-ejection} mechanism could cause the strong outflow of matter necessary to power the GQC oscillations via tidal dissipation. Future outbursts of the source will be crucial to further investigate between the two main mechanisms described above.

\section*{Acknowledgments}
We gratefully acknowledge the Sardinia Regional Government for the financial support (P. O. R. Sardegna F.S.E. Operational Programme of the Autonomous Region of Sardinia, European Social Fund 2007-2013 - Axis IV Human Resources, Objective l.3, Line of Activity l.3.1). This work was partially supported by the Regione Autonoma della Sardegna through POR-FSE Sardegna 2007- 2013, L.R. 7/2007, Progetti di Ricerca di Base e Orientata, Project N. CRP-60529. We also acknowledge financial contribution from the agreement ASI-INAF I/037/12/0. AP acknowledges funding from the European Union's Horizon 2020 research and innovation programme under the Marie Sk\l{}odowska-Curie grant agreement 660657-TMSP-H2020-MSCA-IF-2014, as well as the International Space Science Institute (ISSIBern) which funded and hosted the international team ``The disk magnetosphere interaction around transitional millisecond pulsar''.

\bibliographystyle{mn2e}
\bibliography{biblio}

\label{lastpage}

\end{document}